\documentclass{article}

\usepackage{graphicx}

\begin{document}
\title{Measures of Anisotropy and the Universal Properties of Turbulence}
\author{Susan Kurien\\Physics Department, Yale University, New Haven, CT 06520
\and Katepalli R. Sreenivasan\\Mason Laboratory, Yale University,
New Haven, CT 06520}

\maketitle
\begin{abstract}
Local isotropy, or the statistical isotropy of small scales, is
one of the basic assumptions underlying Kolmogorov's theory of
universality of small-scale turbulent motion. The literature is
replete with studies purporting to examine its validity and
limitations. While, until the mid-seventies or so, local isotropy
was accepted as a plausible approximation at high enough Reynolds
numbers, various empirical observations that have accumulated
since then suggest that local isotropy may not obtain at any
Reynolds number. This throws doubt on the existence of universal
aspects of turbulence. Part of the problem in refining this loose
statement is the absence until now of serious efforts to separate
the isotropic component of any statistical object from its
anisotropic components. These notes examine in some detail the
isotropic and anisotropic contributions to structure functions by
considering their SO(3) decomposition. After an initial discussion
of the status of local isotropy (section 1) and the theoretical
background for the SO(3) decomposition (section 2), we provide an
account of the experimental data (section 3) and their analysis
(sections 4, 5 and 6). Viewed in terms of the relative importance
of the isotropic part to the anisotropic parts of structure
functions, the basic conclusion is that the isotropic part
dominates the small scales at least up to order 6. This follows
from the fact that, at least up to that order, there exists a
hierarchy of increasingly larger power-law exponents,
corresponding to increasingly higher-order anisotropic sectors of
the SO(3) decomposition. The numerical values of the exponents
deduced from experiment suggest that the anisotropic parts in each
order roll off less sharply than previously thought by dimensional
considerations, but they do so nevertheless.
\end{abstract}
\section{Introduction}
Local isotropy, or the isotropy of small scales of turbulent
motion, is one of the assumptions at the core of the belief that
small scales attain some semblance of universality \cite{kolm}.
This is a statistical concept, and is not necessarily opposed to
the idea of structured geometry of small scales \cite{jime}. The
important question about local isotropy is not whether the small
scales are strictly isotropic, but the degree to which the notion
becomes a better approximation as the scales become smaller 
\cite{nelkin}. 
Aside from the generic requirement that the flow Reynolds number be high
(so that small scales as a distinct range may exist independent of
the large scale), the question of how or if local isotropy
becomes a good working approximation depends on the nature of
large-scale anisotropy, on whether or not there are other body
forces, on the nearness to physical boundaries, etc. All of this
was well understood by the time of publication of Monin \& Yaglom
\cite{MY}. The general consensus at the time seems to have been
that small scales indeed attain isotropy far from the boundary at
high enough Reynolds numbers, at least when one considered
second-order quantities.

The situation changed perceptibly when the small scales of scalar
fluctuations were experimentally found to be anisotropic in many
types of shear flows even at the highest Reynolds numbers of
measurement. For a summary, see Ref.~\cite{krs}. Since that time,
various other pieces of evidence are slowly accumulating to
suggest that small-scale velocity is also anisotropic; see, for
example, Ref.~\cite{garg}. The claim is that the previously held
belief---which, to some degree, was comforting---loomed large
only because we had not explored the right statistical quantities.
The notion that anisotropy persists at all scales at all Reynolds
numbers (though manifested only in certain statistical parameters)
puts a strong damper on any theory that purports to consider
small-scale turbulence as a universal object. This is somewhat of
an impasse.
\begin{figure}
\centerline{\includegraphics[scale=0.55]{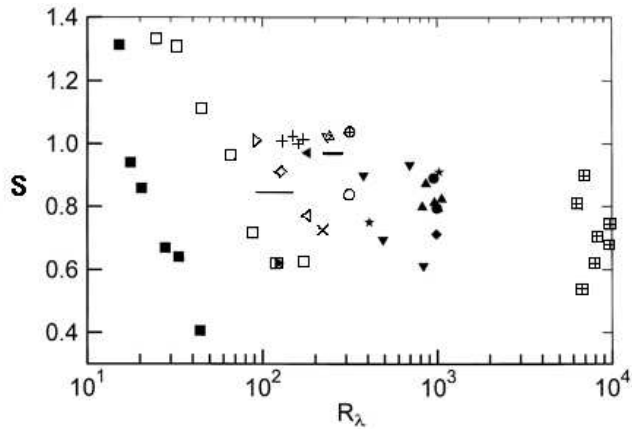}}
\caption{The magnitude of the skewness of the temperature
derivative in turbulent shear flows. The figure is taken from 
Ref.~\cite{sree-anto}. The Taylor microscale Reynolds number $R_{\lambda}$
is proportional to the square root of the large scale Reynolds number. 
It is defined more precisely in section 2.} 
\label{krs_fig}
\end{figure}

Since the problems began with passive scalars, we might as well
consider the evidence in that case a little more closely. The
evidence, collected over many years by many people, is reproduced
in Fig.~\ref{krs_fig} from Ref.~\cite{sree-anto}. The argument is
that if the temperature fluctuation $\theta$ is locally isotropic,
its derivative $\partial \theta/\partial x$, being a small-scale
quantity by construction, should be isotropic. Taking reflection
symmetry as part of isotropy, we should have $\langle (\partial
\theta/\partial x)^n \rangle$ = $-\langle (\partial
\theta/\partial x)^n \rangle$ for all odd values of $n$. This
means that all the odd moments must be zero. In particular,
$\langle (\partial \theta/\partial x)^3 \rangle = 0$, or `small'
in practice. But small compared to what? The standard thing to do
is to normalize $\langle (\partial \theta/\partial x)^3 \rangle$
by $\langle (\partial \theta/\partial x)^2 \rangle^{3/2}$, or to
examine the behavior of the skewness $S$ of $\partial
\theta/\partial x$. These are the data shown in
Fig.~\ref{krs_fig}. The data suggest, despite some large scatter,
that the skewness remains to be of the order unity even at the
largest Reynolds numbers for which measurements are available.

There are two points to be made. The first is that the Taylor
microscale Reynolds number $R_{\lambda}$ may not be the right
parameter against which to plot $S$. As articulated by Hill
\cite{hill}, the reasons are that the velocity that appears in
$R_{\lambda}$ is a large-scale quantity and the length scale
$\lambda$ cannot be defined independent of the large scale
velocity. (In practice, the definition of $\lambda$ may not
require the large scale, see \cite {sree2}, though this is not a
result that can be shown formally; but the comment on the velocity
remains valid in any case.) We believe, however, that the
situation will probably not change qualitatively even if we
adopted a different abscissae for Fig.~\ref{krs_fig}. The second
point is that, if one used $\langle (\partial \theta/\partial x)^4
\rangle^{3/4}$ instead of $\langle (\partial \theta/\partial x)^2
\rangle^{3/2}$ for normalizing $\langle (\partial \theta/\partial
x)^3 \rangle$, the normalized quantity will vanish with Reynolds
number, roughly according to some power of the Reynolds number.
Why is it not legitimate to compare the third moment to the next
high-order even moment, instead of the neighboring low-order
moment, or, perhaps to the geometric mean of the lower and higher
order even moments? In either case, we will have a quantity that
diminishes with the Reynolds number.

These are not elegant arguments, and seem like desperate efforts
made to save the situation at any cost. There is a further
argument to be made, however. Leaving aside the technicalities for
a moment, let us suppose that, for a given statistical quantity to
be measured, there are at all scales an isotropic part and an
anisotropic part. We might now ask whether the ratio of the
anisotropic part to the isotropic part vanishes as the scale size
becomes smaller. We are no longer asking if the third moment
vanishes with respect to (a suitable power of) some even order
moment, but if, within a moment of a given order, the isotropic
part eventually dominates the anisotropic part as the scale size
vanishes. If this is indeed so, it allows us to say that, while
non-universal anisotropic parts may always be present, the
isotropic part dominates at small enough scales. If, of course,
the isotropic part contributes exactly zero to the statistical
quantity being considered, the anisotropic part will no doubt
prevail at any finite Reynolds number, but this is not in contradiction to 
the statement just
made. This is a richer point of view to take, potentially less
inhibiting for the development of a sensible theory of turbulence.
In such a picture, the universal theory that may emerge holds for
the isotropic part alone, but, for it to be applicable to other
types of turbulence, the anisotropic `correction' has to be added
in some suitable way.

The questions of interest, then, are obvious: What is a good way 
to decompose any
statistical object of choice into isotropic and anisotropic parts?
How do these two parts vary with scale size, relative to each
other? A plausible method for answering these questions was
proposed in Ref.~\cite{99ALP} by using the SO(3) decomposition of
tensorial objects usually considered in turbulence. One can argue
as to whether this is the best perspective, but there is no doubt
that it provides one framework within which our questions can be
posed and answered. An experimental assessment of this issue is
the broad topic of these notes. This is preceded by a detailed description
of the theoretical issues involved. 

The notes were part of the lectures presented by KRS at Les Houches,
and form part of the Ph.D. work of SK. We note in advance that the
literature cited is limited to a few key articles.
We thank Itamar Procaccia and Victor L'vov for introducing us to
the subject discussed here, and Christopher White and Brindesh
Dhruva for their help in acquiring some of the data. KRS thanks
the organizers of the School for the invitation to deliver the
lectures.
\section{Theoretical tools}
\subsection{The method of SO(3) decomposition}
A familiar example of decomposition into the irreducible
representations of the SO(3) symmetry group is the solution of the
Laplace equation in spherical coordinates
\begin{equation}
\nabla^2 {\psi}({\bf r}) = 0,\label{lpeq}
\end{equation}
where ${\psi}$ is a scalar function defined over the sphere of radius
$r$. This equation is linear, homogeneous and isotropic. The solutions
are separable, and may be written as
products of functions of $r$, $\theta$ and $\phi$. The solutions form a
linear space, a possible basis for which are derived
from the spherical harmonics $Y_{lm}(\theta,\phi)$
\begin{equation}
\psi_{lm}({\bf r}) = r^l Y_{lm}(\theta, \phi) = r^l Y_{lm}({\bf{\hat r}})
\label{scalar_basis}
\end{equation}
where the index $l=0,1,2,...,\infty$ denotes the degree of the
harmonic polynomial. The angular dependence ($\theta,\phi$) is
equivalent to the unit vector ${\bf{\hat r}}$, or the unit sphere.
In the angular dependence, the index $m$ denotes the elements of
the orthonormal basis space that span each harmonic polynomial of
degree $l$. For each $l$ there are $2l+1$ elements indexed by
$m=-j,-j+1,\dots,j$ which are the $Y_{lm}$. Equation
(\ref{scalar_basis}) is a useful basis space of solutions to the
Laplace equation because the $\psi_{lm}({\bf{\hat r}})$ are each
solutions of the Laplace equation and are orthonormal for
different $l$,$m$. Each basis element has a definite behavior
under rotations, that is, the action on it by an element of SO(3)
preserves the indices. In other words, the rotated basis element
will also be indexed by the same $l$,$m$. The general solution of
(\ref{lpeq}) is given by
\begin{equation}
\psi({\bf r}) = \sum_{l,m} a_{lm}\psi_{lm}({\bf r}).
\label{psi_solution}
\end{equation}
The coefficients $a_{lm}$ are obtained from the boundary conditions on
$\psi({\bf r})$ and, for a particular solution, any of them may be
zero. In the theory of group representations, $\psi_{lm}({\bf r})$ is
said to be the $2l+1$ dimensional {\it irreducible representation} of
the group of all rotations, the SO(3) symmetry group, in the space of
scalar functions over the sphere. The space of scalar functions is called
a `carrier' or `target' space for the SO(3) group representations.
For each irreducible representation
indexed by $l=0...\infty$ there are $2l+1$ components indexed by
$m=-l...+l$. A representation of a symmetry group is said to
be irreducible if it does not contain any subspaces that are invariant
under the transformation associated with that symmetry.

Here we make a distinction between the behaviors under rotation of
the {\it equation} (\ref{lpeq}) and of its {\it solutions}
$\psi({\bf r})$. The statement that the equation is isotropic
means that it will hold true in a rotated frame with the
Laplacian operator ${\overline \nabla}$ and the function
${\overline \psi}({\overline {\bf r}})$ properly defined in the
new coordinate system, ${\overline \nabla}^2 {\overline
\psi}({\overline {\bf r}}) = 0$. However, the function $\psi({\bf
r})$ may contain both isotropic and anisotropic components. All
symmetry groups possess a one-dimensional representation in the
carrier space, which is invariant under the transformations of
that group. For the representations of the SO(3)
symmetry group in the space of scalar functions over the unit
sphere, the one-dimensional representation, indexed by $l=0$, is
$Y_{00} = constant$; it will clearly remain unchanged under proper
rotations. This is the {\it isotropic} representation. The
higher-dimensional representations $Y_{lm}$ (dimension
$3,5,7,\dots$ corresponding to $l=1,2,3,\dots$) have functional
forms in $\theta$ and $\phi$ which are {\it altered} under
rotation even though the degree ($l$), and hence the
dimensionality, of the representation is preserved. These are the
{\it anisotropic} representations. We examine this point in
more detail with further discussion of the simple example of the
scalar functions $\psi$.

For a particular rotation $\Lambda$ in Euclidean space which tells us
how to rotate a vector ${\bf x}$ into a new coordinate system where it is
denoted by ${\bf {\overline x}}$,
\begin{equation}
{\overline x}^\alpha = \Lambda^{\alpha}_{\beta}x^\beta,
\end{equation}
we can define an operator $O_\Lambda$ which tells us how to rotate the
function $\psi({\bf r})$ to ${\overline \psi}({\bf {\overline r}})$.
If the function $\psi({\bf r})$ is written in terms of a linear
combination of its basis elements as in Eq. (\ref{psi_solution}), then
the rotation operation is written as
\begin{eqnarray}
{\overline\psi}({\bf {\overline r}})&=& O_\Lambda\psi({\bf r})\nonumber\\
&=& O_\Lambda\sum_{l,m}a_{lm}\psi_{lm}({\bf r}) \nonumber\\
&=& \sum_{l,m}O_\Lambda a_{lm}\psi_{lm}({\bf r}).
\end{eqnarray}
The $O_\Lambda$ for each $l$ representation is a $(2l+1)\times(2l+1)$
matrix denoted by $D^{(l)}_{m',m}(\Lambda)$. The transformation is
written as
\begin{eqnarray}
O_\Lambda\psi_{lm}({\bf  r})
&=&\sum_{m'=-l}^{l}D^{(l)}_{m',m}(\Lambda) \psi_{lm'}({\bf r}).
\label{Dlm}
\end{eqnarray}
Thus, when a rotation of the function $\psi({\bf r})$ into a new
coordinate system is to be performed, the rotation matrices
$D^{(j)}_{m,m'}(\Lambda)$ are all that is needed in order to transform
each of the irreducible representations of the SO(3) group that form
the basis functions of $\psi({\bf r})$. As can be seen from
Eq.~(\ref{Dlm}), the $j=0$ (one-dimensional) irreducible
representation is the one which is {\it invariant} to all rotations in
that its functional form is always preserved.  The $D^{(0)}(\Lambda)$
matrix is merely a number, independent of $\Lambda$, multiplying the
$j=0$ representation. The $j=0$ irreducible representation is the {\it
isotropic} component of the SO(3) decomposition. For all higher-order
$j$'s the rotation preserves the dimension of the transformed
component, i.e. the indices $j$ and $m$ are retained, but the
functional form is altered. The $D^{(j)}(\Lambda)$ for $j>0$ are true
$(2j+1)\times(2j+1)$ matrices that mix the various $m$ contributions
of the original, unrotated basis.

The above simple example involved the case of a scalar function
$\psi$ that depended on a single unit vector ${\bf {\hat r}}$. The
method of SO(3) decomposition may be carried over to more
complicated objects. The theory is given in detail in
\cite{99ALP}.  In general we can imagine an $n^{th}$ order tensor
function which depends on $p$ unit vectors
$T^{\alpha_1\alpha_2\dots\alpha_n}({\bf{\hat r_1}}, {\bf{\hat
r_2}},\dots {\bf{\hat r_p}})$ and is the solution to an isotropic
equation. Then, the rules of SO(3) decomposition carry over in the
following manner. The rotation operator $O_\Lambda$ is now 
defined through the relation
\begin{eqnarray}
&&{\overline T}^{\alpha_1\alpha_2\dots\alpha_n}
({\overline {\bf {\hat r_1}}},{\overline {\bf {\hat r_2}}},...
{\overline {\bf {\hat r_p}}})= O_\Lambda
T^{\alpha_1\alpha_2\dots\alpha_n}({\bf {\hat r_1}},{\bf {\hat r_2}},\dots,
{\bf {\hat r_p}}) \nonumber\\
&=& \Lambda^{\alpha_1}_{\beta_1}\Lambda^{\alpha_2}_{\beta_2}\dots
\Lambda^{\alpha_n}_{\beta_n} T^{\beta_1\beta_2\dots\beta_n}
(\Lambda^{-1}{\overline{\bf {\hat r_1}}},\Lambda^{-1}
{\overline{\bf {\hat r_2}}},\dots,\Lambda^{-1}{\overline{\bf{\hat r_p}}}).
\end{eqnarray}
The tensor function $T$ may be written in terms of the irreducible
representations of the SO(3) symmetry group. The basis elements
are denoted by $B^{\alpha_1\alpha_2\dots\alpha_n}_{jm}({\bf {\hat
r}})$ where the $j$ index plays the same role as the $l$ index in
the example of the scalar function. The basis elements are more
complicated than the $\psi_{lm}$ because the space of functions
$T$ is the direct product of $n$ Euclidean three-dimensional
vector spaces (manifest in the indices $\alpha_1\dots\alpha_n$)
with $p$ infinite dimensional spaces of continuous single-variable
functions over the unit sphere (the ${\bf {\hat r_1}}\dots{\bf
{\hat r_p}}$). If the constituent spaces are also written in the
SO(3) decomposition, the rules of angular momentum addition,
familiar from quantum mechanics, may be used in taking the direct
product. The three-dimensional Euclidean vector space is a $j=1$
space while each of the infinite-dimensional spaces functions over
the unit sphere is the sum of the $j=0, j=1, \dots,
j\rightarrow\infty$ irreducible representations of SO(3) (recall
the above example of $\psi({\bf {\hat r}})$) with each $j$
representation appearing once.

In tensor product notation, the
product space of two vector spaces $V_1$ and $V_2$ is denoted by
$V_1 \otimes V_2$. In the case of vector spaces in the SO(3) notation 
the spaces are named uniquely by their index $j$. Using tensor product 
notation, the $n$ three-dimensional Euclidean spaces form a space
$1\otimes1\otimes\dots\otimes1$ ($n$ times). The tensor notation
indicating a linear sum of tensors with SO(3) representation
indices $j_1$ and $j_2$ is $j_1\oplus j_2$. Using this notation,
each of the infinite dimensional spaces is $0 \oplus 1\oplus
2\oplus \dots$. The direct product of $p$ of these is ($0 \oplus
1\oplus 2\oplus \dots) \otimes (0 \oplus 1\oplus 2\oplus
\dots)\otimes \dots \otimes(0 \oplus 1\oplus 2\oplus \dots)$ ($p$
times). Thus, the final product space for $B_{jm}({\bf {\hat r}})$
is written in tensor-product notation as
$1\otimes1\otimes\dots\otimes1$ ($n$ times) $\otimes (0 \oplus
1\oplus 2\oplus \dots) \otimes (0 \oplus 1\oplus 2\oplus
\dots)\dots (0 \oplus 1\oplus 2\oplus \dots)$ ($p$ times). Now,
the tensor product of spaces $j_1$ and $j_2$ will contain new
spaces whose SO(3) indices are given in the following manner. We
recall the rules of angular momentum addition familiar from
quantum mechanics. The total angular momentum of an SO(3)
representation space is given by its index $j$. The rules of
angular momentum addition dictate how the product of two spaces of
angular momentum (SO(3) index) $j_1$ and $j_2$ may be added and
the possible $j$ indices of the resulting spaces,
\begin{equation}
j_1 \otimes j_2 = |j_1 - j_2|\oplus \dots \oplus (j_1 + j_2).
\label{addrule}
\end{equation}
Equation (\ref{addrule}) says that the direct product of two spaces
each belonging to a particular $j$ will generate a sum of new spaces
with only those $j$ indices allowed by the rule. The operations
$\otimes$ and $\oplus$ are distributive like the corresponding arithmetic
operators. For example
\begin{eqnarray}
1\otimes 1 &=& 0\oplus1\oplus 2\nonumber\\
1\otimes 1 \otimes 1 &=& (0\oplus1\oplus2)\otimes1\nonumber \\
&=& (0\otimes1) \oplus (1\otimes1)\oplus (2\otimes 1).
\label{1c1}
\end{eqnarray}
If we now apply the angular momentum addition rules
(\ref{addrule}) to each of the product terms on the right hand
side of Eq.~(\ref{1c1}), we get
\begin{eqnarray}
1\otimes 1 \otimes 1&=& 1\oplus 0 \oplus 1 \oplus 2 \oplus 1 \oplus 2
\oplus 3 \nonumber \\
&=& 0 \oplus (1 \times 3)\oplus (2 \times 2) \oplus 3 .
\label{1c1c1}
\end{eqnarray}
When taking the product of more
than two spaces, there will be several ways to arrive at a particular
$j$ in the final product space. We see this in the
final rearrangement of the right hand side of Eq.~(\ref{1c1c1}),
where there is only one representation each of $j=0$ and $j=3$, but
{\it three} of $j=1$ and {\it two} of $j=2$. Going back to our
more complicated target space for the basis elements
$B^{\alpha_1\alpha_2\dots \alpha_n}_{jm}({\bf {\hat r}})$,
which is $1\otimes1\otimes\dots\otimes1$ ($n$ times) $\otimes
(0 \oplus 1\oplus 2\oplus \dots) \otimes
(0 \oplus 1\oplus 2\oplus \dots)\dots (0 \oplus 1\oplus 2\oplus \dots)$
($p$ times),
we expect that there may be many ways to obtain a
particular $j$ in the final product space. To indicate this, for each
basis element we associate a further index $q$ giving
$B_{qjm}({\bf {\hat r}})$. These are important when we start to
actually calculate the basis elements. The machinery used is the
Clebsch-Gordan method well-known in quantum mechanics and the reader is
referred to \cite{99ALP} for the details. The tensor $T$ may also
depend on {\it vectors} ${\bf r_i}$ with
magnitudes $r_i$ but for each rotation operation $O_\Lambda$ these may be
treated as parameters included in the weight associated with each basis
element. The $T$ are represented as
\begin{eqnarray}
T^{\alpha_1\alpha_2\dots\alpha_n}({\bf r_1},{\bf r_2},\dots,{\bf r_p}) =
\sum_{q,j,m}a_{qjm}(r_1,r_2,\dots,r_p)\nonumber\\
\times B^{\alpha_1\alpha_2\dots\alpha_n}_{qjm}({\bf{\hat r_1}},
{\bf{\hat r_2}},\dots,{\bf{\hat r_p}}).
\label{T_SO3}
\end{eqnarray}

In the next two sections, we move away from the very general formalism of
SO(3) decomposition discussed above, and apply it to the specific case
of the statistical tensor quantities in fluid turbulence and
their dynamical equations obtained from the Navier-Stokes equations.
We are motivated by the observation that the SO(3) decomposition when
applied in the correct manner would allow ($a$) separation of the
scaling variable $r$ from the angular dependence, and ($b$) the separation
of the isotropic from the anisotropic parts of the tensor.

\subsection{Foliation of the structure function into $j$-sectors}

We would like to use the formalism of the SO(3) representation, as
presented in general terms in the previous section, to study
the structure function tensor in turbulence theory. The $n$th order
structure function is defined by an ensemble average of the $n$th moment
of the difference of velocity components across scales $r_p$. It is in
fact a tensor function over $p$ unit spheres. We define the
velocity difference as
\begin{equation}
w^{\alpha_i}({\bf r_k}) = u^{\alpha_i}({\bf x+r_k})-u^{\alpha_i}({\bf x}),
\label{w}
\end{equation}
where the $\alpha_i$ denotes the component of the velocity
vector in the direction $i$ in a defined coordinate system, and the
subscript $k$ on ${\bf r_k}$ denotes a particular choice of the vector.
The $n$th order structure function for $p$ such choices of vector scale
${\bf r}$ is then
\begin{equation}
S^{\alpha_1\alpha_2\dots\alpha_n}({\bf r_1},{\bf r_2},\dots,{\bf r_p})
= \langle w^{\alpha_1}({\bf r_i})w^{\alpha_2}({\bf r_j})\dots
\times w^{\alpha_n}({\bf r_k})\rangle,
\label{Sn}
\end{equation}
where the subscripts $i$,$j$,$\dots$ denote any of the vectors
from ${\bf r_1}$ to ${\bf r_p}$.

In this section we review briefly the reason why the structure
functions may be written as a linear combination of the
irreducible representations of the SO(3) symmetry group. For
details the reader is again referred to \cite{99ALP}.  The
dynamical equation for the structure function may be derived from
the Navier-Stokes equations. Similar to the solutions of the
Laplace equation, the solution to the $n${th} order structure
function equation forms a linear space with the basis chosen
to be the irreducible representations $B_{qjm}({\bf r})$ of the
SO(3) group as shown in previous sections (see Eq.~(\ref{T_SO3}).
Since the basis is orthonormal for different $j,m$, and the
equation is isotropic, we obtain a $heirarchy$ of dynamical
equations each with terms of a $given$ $j$ and $m$. This
demonstrates that the dynamical equations themselves $do$ $not$
mix the various $j$ and $m$ contributions, and that the dynamical
equation for the $isotropic$ part of the structure function is
different from the dynamical equations for any of the higher $j$
contributions. Further, it may be shown from solvability
conditions on this set of equations that the scaling part of the
function changes between $j$-sectors while remaining the same
within a $j$-sector, for different $m$. This motivates the
postulate that the different $j$ sectors scale with exponents
different from the $isotropic$ $\zeta_n$. We denote these by
$\zeta^{(j)}_n$ where subscript $n$ indicates the rank of the
tensor and superscript $(j)$ indicates the number of the
irreducible representation. We will leave out the superscript in
the case of $j=0$ since then we recover the known isotropic
scaling exponent $\zeta_n$.

\subsection{The velocity structure functions}

In the previous section we provided a heuristic justification for
the use of the SO(3) decomposition for this tensor using the fact
that it is the solution of an isotropic equation. Now, we use the
rules of SO(3) (see Eq.~(\ref{T_SO3})) decomposition to write it
as
\begin{eqnarray}
S^{\alpha_1\alpha_2\dots\alpha_n}({\bf r_1},{\bf r_2},\dots,{\bf r_p})
&=& \sum_{q,j,m}a_{qjm}(r_1,r_2,\dots,r_p)\nonumber\\
&\times& B^{\alpha_1\alpha_2\dots\alpha_n}_{qjm}({\bf {\hat r_1}},
{\bf {\hat r_2}},\dots,{\bf {\hat r_p}}).
\label{Sn_SO3}
\end{eqnarray}

In this form, as was demonstrated for the scalar
function $\psi$, the $j=0$ representation is the isotropic one.
Its prefactor $a_{q_{00}}(r_1,r_2,\dots,r_p)$ contains the scale
dependence. For $p>1$ there will be infinitely many ways to obtain
a particular $j$ for the product space as may be seen from the
angular momentum addition rules in Eq.~(\ref{addrule}) and the
direct-product representation. Therefore, $q$ ranges from $1$ to
$\infty$. This is a difficult hurdle computationally but,
fortunately, most experimental and theoretical work deals with
the dependence on a single vector, and $p=1$. In that case, and for
rank $n=2$, we obtain the second order structure function
$S^{\alpha\beta}({\bf r})$ quite easily.

\subsubsection{The second-order structure function}

We consider the structure function of Eq.~(\ref{Sn_SO3}) for $n=2$
and $p=1$. The tensor product space is
\begin{eqnarray}
1\otimes 1\otimes (0\oplus 1\oplus 2\oplus \dots)&=&(0\oplus1\oplus2)
\otimes(0\oplus 1\oplus 2\oplus \dots) \nonumber\\
&=& (0 \otimes 0)\oplus(0 \otimes 1)\oplus(0 \otimes 2)\oplus\dots
\nonumber\\
&\oplus& (1 \otimes 0)\oplus(1 \otimes 1)\oplus(1 \otimes 2)\oplus \dots
\nonumber\\
&\oplus& (2 \otimes 0)\oplus(2 \otimes 1)\oplus(2 \otimes 2)\oplus \dots.
\end{eqnarray}
As demonstrated in (\ref{1c1c1}) using the addition rule
(\ref{addrule}), there could be more than one way of obtaining a 
particular $j$ representation is obtained in the product space. To count
these for a given j, we had to add there the index $q$ to the basis
tensors (see for example Eq.~(\ref{T_SO3})). We find for the
second-order tensor over a single sphere that
\begin{itemize}
\item $j=0$ has a total of 3 representations
\item $j=1$ has a total of 7 representations
\item $j>1$ has a total of 9 representations.
\end{itemize}

The Clebsch-Gordon machinery tells us in addition about the symmetry 
(in the indices) and parity (in $r$) of each $q$ contribution. Using
this information, the terms of the basis may be constructed. As in
the case of the scalar function, the $B^{\alpha\beta}_{qjm}$ are
orthonormal for different $j$,$m$ and $q$. In practice, the
Clebsch-Gordon method of constructing these objects is rather
tedious and so we follow the alternative method offered in
\cite{99ALP}. We make use of the Clebsch-Gordon methods only to
obtain the number of representations in each $j$, and its parity
and symmetry properties. Armed with this information, the
$B^{\alpha\beta}_{qjm}$ may be constructed by a more convenient
means. The idea is to use the fact that we already know an
orthonormal basis in the SO(3) representation for the scalar
function over the sphere (Eq.~(\ref{scalar_basis})). We can now
`add indices' to this basis in a way to be described, while
retaining the properties under rotation, in other words its $j$
and $m$ values. To add indices in a way that does not change the
$j$,$m$ values is to perform contraction with the objects
$\delta^{\alpha\beta}$,$r^{\alpha}$,$\epsilon^{\alpha\beta\gamma}$
and the partial derivative operator $\partial_\alpha$. This method
automatically takes care of the $j$ and $m$ properties; and all
that needs to be done now is to apply the above operators to obtain
the different $q$ terms with the right symmetry and parity
properties. This gives us a complete set of basis
elements which are different than the Clebsch-Gordan method. 
The orthonormality among
different $q$ for the same $j$ and $m$ is lost, but the
orthonormality among different $j$ and $m$ is maintained because
we start out with a basis Eq.~(\ref{scalar_basis}) which already
possesses these properties. However, different $q$ elements are
linearly independent and span  a given ${j,m}$-sector. The details
on how this is done using the rules from \cite{99ALP} is
presented in appendix A. In what follows we simply write down the
components calculated using that method.

The second order structure function tensor is
\begin{equation}
S^{\alpha\beta}({\bf r}) = \langle (u^\alpha({\bf x+r})
- u^\alpha({\bf x}))(u^\beta({\bf x+r})- u^\beta({\bf x}))\rangle
\label{S2}
\end{equation}
which we decompose using the SO(3) irreducible representations
$B^{\alpha\beta}_{qjm}({\bf {\hat r}})$ as
\begin{eqnarray}
S^{\alpha\beta}({\bf r}) &=& S^{\alpha\beta}_{j=0}({\bf r}) +
S^{\alpha\beta}_{j=1}({\bf r})+S^{\alpha\beta}_{j=2}({\bf r})+\dots
\nonumber\\
&=& \sum_{q,j,m}a_{qjm}(r) B^{\alpha\beta}_{qjm}(\alpha\beta)
({\bf {\hat r}}).
\label{S2decomp}
\end{eqnarray}
A further constraint is provided by the incompressibility condition
\begin{equation}
\partial_\alpha S^{\alpha\beta}({\bf r}) = 0.
\end{equation}
Each $j$,$m$-sector must separately satisfy the incompressibility
condition since taking the partial derivative preserves the
rotation properties and does not mix the $j,m$ sectors. Therefore,
the incompressibility condition provides a constraint among the
different $q$ contributions within a given $j,m$-sector. For
example, the $j=0$ contribution has three different
representations ($q=\{1,2,3\}$), two of which are symmetric and
one antisymmetric in the indices $\alpha$, $\beta$. By
definition, the structure function is symmetric in the indices, 
therefore the antisymmetric contribution will not appear giving only 
two $q$ contributions. The incompressibility constraint gives
\begin{eqnarray}
\partial_\alpha\sum_q a_{q_{00}}r^{\zeta_2}B^{\alpha\beta}_{q_{00}}
({\bf {\hat r}})= 0, \nonumber \\
\end{eqnarray}
and
\begin{eqnarray}
\partial_\alpha(a_{100}r^{\zeta_2}\delta^{\alpha\beta}+
a_{200}r^{\zeta_2}{r^\alpha r^\beta \over r^2}) = 0.
\label{incomp}
\end{eqnarray}
Motivated by the expectation that the structure functions scale as
powers of $r$ in the inertial range, we assumed in Eq.~(\ref{incomp})
that the scale-dependent prefactor is of the form
$a_{q_{00}}r^{\zeta_n}$ where $a_{q_{00}}$ is a flow-dependent
constant. Equation~(\ref{incomp}) results in a relationship
between $a_{100}$ and $a_{200}$ giving the final form of the isotropic
$j=0$ sector with just one unknown coefficient $c_0$ to be determined
by the flow boundary conditions. That is,
\begin{equation}
S^{\alpha\beta}_{j=0}({\bf r})=
c_0r^{\zeta_2}\Big[(2+\zeta_2)\delta^{\alpha\beta} -
\zeta_2{r^\alpha r^\beta \over r^2}\Big].
\label{S2_incomp}
\end{equation}
Here, $\zeta_2\approx0.69$ is the known empirically known
anomalous second-order scaling exponent. We would like to assume a
similar scaling form for the prefactor $a_{qjm}(r)$ for $j>1$. In
such a formulation, there is a hierarchy of scaling exponents
which we denote by $\zeta^{(j)}_2 \neq \zeta_2$ corresponding to
the higher-order $j$ sectors. Successive $j$'s indicate increasing
degrees of anisotropy. The following section provides a
justification for such a classification of the scaling of the
various sectors. It is the larger goal of this article to examine
these aspects of the theory with the help of high-Reynolds-number
data.

\subsection{Dimensional estimates for the lowest order anisotropic
scaling exponents}

In this section we present dimensional considerations to determine
the ``classical'' values expected for $\zeta_2^{(1)}$ and
$\zeta_2^{(2)}$  the spirit of Kolmogorov's 1941 theory
(henceforth called K41).  We work on the level of K41 to produce
the value $\zeta_2^{(0)}=2/3$. Ignoring intermittency corrections
is justified to the lowest order because the differences between any
two values $\zeta_2^{(j)}$ and $\zeta_2^{(j')}$ for $j\ne j'$ are
considerably larger than the intermittency corrections to either
of them.

It is easiest to produce a dimensional estimate for $\zeta_2^{(2)}$.
One simply asserts \cite{lumley} that the $j=2$ contribution is the
first one appearing in $S^{\alpha\beta}({\bf r})$ due to the existence
of a shear. Since the shear is a second rank tensor, it can appear
linearly in the $j=2$ contribution to $S^{\alpha\beta}({\bf r})$. We
thus have for any $m$, $-j\le m\le j$,
\begin{equation}
S_{j=2}^{\alpha\beta}({\bf r}) \sim T^{\alpha\beta\gamma\delta}
{\partial \bar U^\gamma\over \partial r^\delta} f(r,\bar \epsilon). 
\label{Sshear}
\end{equation}
Here $T^{\alpha\beta\gamma\delta}$ is a constant dimensionless
tensor made of $\delta^{\alpha\beta}$, $\frac{r^\alpha}{r}$, and
{\em bilinear} contributions made of the three unit vectors ${\hat
{\bf p}},~{\hat{\bf m}},~{\hat {\bf n}}$, which form our
coordinate system as defined in appendix A. The form of
Eq.~(\ref{Sshear}) means that the dimensional function
$f(r,\bar\epsilon)$ stands for the response of the second-order
structure function to a small external shear. Within the K41
dimensional reasoning, this function in the inertial interval can
be made only of the mean energy flux per unit time and mass, $\bar
\epsilon$ and $r$ itself. The only combination of $\bar \epsilon$
and $r$ that yields the right dimensions of the function $f$ is
$\bar \epsilon^{1/3} r^{4/3}$. Therefore
\begin{equation}
S_{j=2}^{\alpha\beta}({\bf r}) \sim T^{\alpha\beta\gamma\delta}
{\partial \bar U^\gamma\over
\partial r^\delta}
\bar \epsilon^{1/3} r^{4/3} \ . \label{yofi1}
\end{equation}
We thus find a ``classical K41" value of $\zeta_2^{(2)}$ to be $4/3$.

To obtain the value of $\zeta_2^{(1)}$ we cannot proceed in the
same way. We need a contribution that is linear (rather than
bilinear) in the unit vectors ${\hat {\bf p}},~{\hat{\bf
m}},~{\hat{\bf n}}$. We cannot construct a contribution that is
linear in the shear and yet does not vanish due to the
incompressibility constraint. Thus there is a fundamental
difference between the $j=2$ term and the $j=1$ term. While the
former can be understood as an inhomogeneous term linear in the
forced shear, the $j=1$ term, being more subtle, may perhaps be
connected to a solution of some homogeneous equation well within
the inertial interval.

We therefore need to consider some quantity other than the shear
which could contribute to the anisotropy.  One invariance in the
inviscid limit is as given by Kelvin's circulation theorem. In the
so-called Clebsch representation, one writes the Euler equation in
terms of one complex field $a({\bf r},t)$, see for
example\cite{91Lvo}.  In the ${\bf k}$-representation, the Fourier
component of the velocity field ${\bf u}({\bf k},t)$ is determined
from a bilinear combination of the complex field
\begin{eqnarray}
{\bf u}({\bf k},t)&=&\frac{1}{8\pi^3}\int d^3k_1 d^3k_2
{\bf \Psi}({\bf k_1},{\bf k_2})a^*({\bf k_1},t)
a({\bf k_2},t)\, \\
{\bf \Psi}({\bf k_1},{\bf k_2})&=&\frac{1}{2}\left({\bf k_1}+{\bf
k_2}- ({\bf k_1}-{\bf k_2}) \frac {k_1^2-k_2^2}{|{\bf k_1}-{\bf
k_2}|^2}\right). 
\end{eqnarray} It was argued in \cite{91Lvo} that
this representation reveals a local conserved integral of motion
given by
\begin{equation}
{\bf \Pi} = \frac{1}{8\pi^3}\int d^3k {\bf k} a^*({\bf k},t)a({\bf
k},t). 
\end{equation} 
Note that this conserved quantity is a
vector, and so it cannot have a finite mean in an isotropic
system. Consider a correction to the second order  structure
function due to a flux of the integral of motion $\bar{\bf \pi}$.
The dimensionality of $\bar{\bf \pi}$ is $[\bar {\bf \pi}]
=[\bar\epsilon^{2/3}/r^{1/3}]$, and therefore now the
dimensionless factor is $\bar {\bf \pi}\,r^{1/3}
/\bar\epsilon^{2/3}$. As we did with the shear in the case of the
$j=2$, we assume here the expandability of $\delta_ {\bf \pi} S$
at small values of the flux $\bar{\bf \pi}$ and find
\begin{equation} S_{j=1}^{\alpha\beta}({\bf r}) \sim
T^{\alpha\beta\gamma}\,\bar\pi^\gamma\, r \, 
\label{yofi2}
\end{equation}
where $T^{\alpha\beta\gamma}$ is a constant dimensionless tensor
that is {\em linear} in the unit vectors $\hat {\bf p},~\hat {\bf
m} ,~\hat{\bf n}$. We thus find the ``classical K41" value
$\zeta_2^{(1)}$ is $1$. The derivation of the exponents
$\zeta_2^{(1)}$ and $\zeta_2^{(2)}$ is given in \cite{KLPS00}. It
is concluded that dimensional analysis predicts values of 2/3, 1
and 4/3 for $\zeta_2^{(j)}$ with $j=0$, 1 and 2 respectively. It
is not known at present how to continue this line of argument for
$j>2$.

For the higher-order structure functions
$S^{\alpha_1 \alpha_2 \dots \alpha_n}({\bf r})$ where $n>2$, similar
dimensional analysis may be performed in order to find the K41
contribution to scaling due to shear (corresponding to the $j=2$
component). For each $n$, the lowest order
correction to scaling that is linear in the shear is $n/3 + 2/3$.

\subsection{Summary}
The technique of SO(3) decomposition may be used in order to write
the structure function tensor in terms of its isotropic part,
indexed by $j=0$, and higher-order anisotropic parts, indexed by
$j>0$. The dynamical equation for the structure function of order
$n$ foliates into a set of equations each of different $j,m$. This
motivates the postulate that the different sectors scale
differently. The theory also indicates that any scaling behavior
would depend only the $j$ index and be independent of the $m$
within that sector. There is allowance for dependence on boundary
conditions, specific kinds of forcing and so on, in the unknown
coefficients $a_{qjm}$ in the SO(3) expansion (Eq.~(\ref{T_SO3})).
We first consider the second-order structure function in a
detailed manner in the light of this group representation. The
analysis may be implemented for structure functions of any order
but the task becomes computationally and conceptually more
difficult for $n>2$. In section 6, we present a means of
circumventing these problems.

We have reviewed the theoretical estimates for the low-order
anisotropic scaling exponents for the second-order structure
function. The method may be carried over to the higher-order
objects as well, and we have derived the lowest order
shear-dependent scaling contributions to the $n^{th}$ order
structure function to be $r^{n/3 + 2/3}$. For the second-order
structure function, we have shown that the $j=2$ contribution
corresponds to a low-order shear-dependency. Again for the
second-order structure function, we have presented a conserved
quantity in the Clebsch-representation which provides the correct
dimensional properties for the $j=1$ scaling contribution to be
$r$. The arguments used are purely dimensional in the K41 sense,
and do not take account of anomalous scaling. We expect that real
turbulent flows will exhibit anomalous scaling in the isotropic
sector, $and$ in every anisotropic sector in the SO(3) hierarchy.
The issue of anomalous exponents in turbulence has now multiplied
several-fold, to all the $j$ sectors, in light of the apparent
universality that this work suggests. Thus, the
theoretically predicted anisotropic exponents are to be treated
merely as estimates, especially for higher-order structure
functions ($n>2$), since the scaling anomalies are expected to 
increase with the order $n$, similar to the isotropic case.

In subsequent sections we demonstrate the use of experimental data
in order to test the predictions of the theory. In particular, we
provide explicit calculations of measurable tensor quantities and
extract scaling corrections due to anisotropy.
\section{Some experimental considerations}
\subsection{Background}
Experimental studies of turbulent flows at very high Reynolds numbers
are usually limited in the sense that one measures the velocity field
at a single spatial point, or a few spatial points, as a function of
time \cite{MY}, and uses
Taylor's hypothesis to identify velocity increments at different times
with those across spatial length scales, $r$. The standard outputs of
such single-point measurements are the longitudinal two-point
differences of the
Eulerian velocity field and their moments. In homogeneous
and isotropic turbulence, these structure functions are observed to vary
as power-laws in $r$, with scaling exponents $\zeta_n$ \cite{Fri}.

Recent progress in measurements and in simulations has begun to
offer information about the tensorial nature of structure
functions. Ideally, one would like to measure the tensorial $n$th
order structure functions defined in Eqs.~(\ref{w}) and
(\ref{Sn}). Such information should be useful in studying the
anisotropic effects induced by all practical means of forcing.

\subsection{Relevance of the anisotropic contributions}
In analyzing experimental data the model of ``homogeneous and
isotropic small-scale" is universally adopted, but it is important
to examine the relevance of this model for realistic flows. As we
will demonstrate in the next chapter, our data we use exhibit
anisotropy down to fairly small scales \cite{sreeni}. We have shown
mathematically that keeping the tensorial information helps
significantly in disentangling different scaling contributions to
structure functions. In the light of the SO(3) representation of
section 2, where it is shown that anisotropy might lead to
different scaling exponents for different tensorial components, a
careful study of the various contributions is needed. This is our
goal in the rest of this article.

\subsection{The measurements}
In order to extract a particular $j$ contribution and the associated
scaling exponent, one would ideally like to possess the statistics of
the velocity at all points in three-dimensional space.  One could
then extract the $j$ contribution of particular interest by
multiplying the full structure function by the appropriate $B_{qjm}$
and integrating over a sphere of radius $r$. Orthogonality of the
basis functions ensures that only the $j$ contribution survives the
integration.  One could then perform this procedure for various $r$
and extract the scaling behaviors.

The method just described was adopted successfully in \cite{99ABMP} 
using data from direct numerical simulations of channel flows. The Taylor
microscale Reynolds number $R_{\lambda}$ for the simulations was
about 70. This is not large enough for a clear inertial scaling
range to exist. The authors of Ref. \cite{99ABMP} resort to
extended self-similarity (ESS) in both the isotropic and the
anisotropic sectors. Nonetheless, the results indicated a
scaling exponent of about $4/3$ in the $j=2$ sector.
While the experimental data are limited to a few points in space,
and the integration over the sphere is not possible, we are able
to attain very high Reynolds numbers especially in the atmosphere
under steady conditions ($R_{\lambda}\approx 10,000-20,000$).
Despite the advantage of extended scaling range, we are however
faced with a true superposition of contributions from various $j$
sectors with no simple way of disentangling them as was done with
the numerical data. However, as we will show in section 6,
we can make a judicious choice of the tensor components studied,
and obtain access to the anisotropic contributions.

The tensor structure of the velocity structure functions is lost
in the computation of the usual single-point single-component
measurement of longitudinal and transverse objects. This is
because the part of the expansion that is dependent on the angle
$\theta$ is hidden: the longitudinal and transverse components set
the value of $\theta$ to a constant. The boundary-dependent
prefactors now collapse to a single number as
\begin{equation}
S^{\alpha\beta}_{j=2}(r) = ar^{\zeta^{(j)}_2}.
\label{j2_singlept}
\end{equation}
On the one hand, this is a simple expression. On the other, the
angular dependence is completely lost and there is no formal
difference from the isotropic object
\begin{equation}
S^{\alpha\beta}_{j=0}(r) = c_0r^{\zeta_2}.
\label{j0_singlept}
\end{equation}

\begin{figure}
\centerline{\includegraphics[scale=0.7]{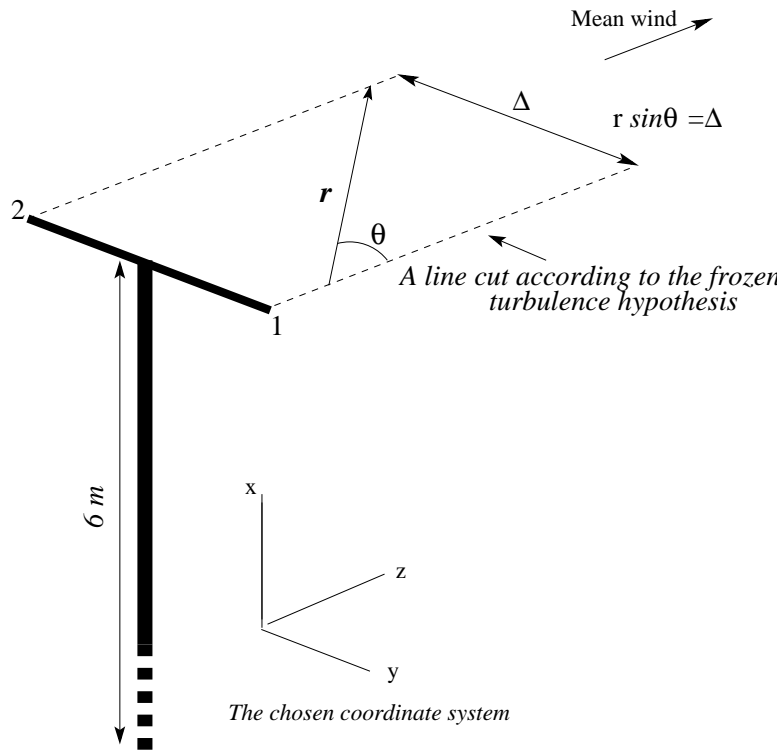}}
\caption{Schematic of the experimental set-up. Shown is the positioning
of the single wire probes 1 and 2 with respect to the mean wind,
and a visual explanation of how Taylor's hypothesis is employed. The coordinate
system chosen is different from what is conventional in turbulence---where
the mean-wind is taken to lie along the $x$-axis. The pictured choice of
coordinates
simplifies the calculations involving the spherical harmonics.}
\label{expsetup}
\end{figure}

In order to see the true tensor character of the structure
function we need an angular variation of the scale separation $r$.
In the atmospheric boundary layer which offers the highest
Reynolds numbers available, the simplest configuration that would
allow us to do this is a two hot-wire combination separated by
distance $\Delta$ in the spanwise direction (y), orthogonal to the
mean-wind. By Taylor's frozen flow hypothesis, such a set-up
will provide two simultaneous one-dimensional cuts through the
flow. Therefore, one can measure the correlation between the two
probes, across a scale $r$ that makes an angle $\theta$ with the
mean-wind direction. As $r$ varies, the angle with respect to the
mean-wind will also vary, giving some functional dependence on
$\theta$ for the coefficients in the SO(3) decomposition. A
schematic of the experimental configuration is presented in Fig.
\ref{expsetup}.

A final consideration in measurements aimed at measuring
anisotropic contributions is the homogeneity of the flow. The
incompressibility condition may only be applied as a constraint on
the structure function coefficients if the flow is homogeneous in
the $r$ direction. If we consider, for instance, probes separated
in the shear (vertical) direction, $r$ will have a component in
the inhomogeneous direction and incompressibility may not be used
to constrain the terms in the S0(3) expansion, leaving too many
free parameters. However, in some instances, it is
important to consider quantities that are not constrained by the
incompressibility condition. For example, the $j=1$ contribution
to the second order structure function will not appear in the
experimental configuration described above. This is so because
only one symmetric, even-parity, term exists in $j=1$ by the
Clebsch-Gordon rules, and this must vanish by the
incompressibility constraint. Probe-separation in the shear
direction would be needed to produce a non-zero contribution of
the $j=1$ term. In short, one can understand the nature of the
best experimental configuration from carefully studying the tensor
decomposition of the structure function.

\begin{table}
\centerline{\begin{tabular} {|c|c|c|c|c|c|c|c|c|}\hline
Height&$\overline U$ & $u^\prime$ &$10^2 \langle \varepsilon \rangle$,&$\eta$
& $\lambda$ & $R_{\lambda}$ & $f_s,$ per & \# of \\meters& ms$^{-1}$ & ms
$^{-1}$
& m $^2$ s$^{-3}$ & mm & cm & & channel, Hz & samples\\
\hline 6 & 4.1 & 1.08 & $1.1$ & 0.75 & 15 & 10,500 & 10,000 & $4 \times
10^7$\\
\hline
35 & 8.3 & 2.30 & $7.8 $ & 0.45 & 13 & 19,500 & 5,000 & $ 4 \times 10^7$\\
\hline
0.11 & 2.7 & 0.47 & $6.6 $ & 0.47 & 2.8 & 900 & 5,000 & $ 8 \times 10^6$\\
0.27 &3.1 & 0.48 & 2.8& 0.6 & 4.4 & 1400& 5,000 &$8 \times 10^6$\\
0.54 &3.5 &0.5& 1.5& 0.7& 6.2&2100& 5,000& $8 \times
10^6$\\
\hline
\end{tabular}}
\caption{\label{exp_data}Data sets I (first line), II (second
line) and III (third-fifth lines). The various symbols have the
following meanings: $\overline U$ = local mean velocity,
$u^{\prime}$ = root-mean-square velocity, $\langle \varepsilon
\rangle$ = energy dissipation rate obtained by the assumption of
local isotropy and Taylor's hypothesis, $\eta$ and $\lambda$ are
the Kolmogorov and Taylor length scales, respectively, the
microscale Reynolds number $R_{\lambda} \equiv u^{\prime}
\lambda/\nu$, and $f_s$ is the sampling frequency.}
\end{table}

We analyze measurements in atmospheric turbulence at various
heights above the ground (data sets I, II and III). Sets I and III
were acquired from flow over a long fetch in the salt flats in
Utah. The site of measurements was chosen to provide steady wind
conditions. The surface of the desert was smooth and even: the
water that floods the land during spring recedes uniformly and
leaves the ground hard and essentially smooth during early summer. The
boundary layer on the desert floor in early summer is thus quite similar 
to that on a smooth flat plate \cite{klew}. The measurements were made in
the early summer season roughly between 6~PM and 9~PM during which
nearly neutral stability conditions prevailed. Data set II was
acquired over a rough terrain with ill-defined fetch at the
meteorological tower at the Brookhaven National Laboratory. In
sets {\rm I} and {\rm II}, data were acquired at heights 6 m and
35 m respectively. They were recorded simultaneously from two
single hot-wire probes separated in the spanwise direction $y$ by
55 cm and 40 cm respectively. In both cases, the separation
distance was within the inertial range, and was set nominally
orthogonal to the mean wind direction (see below). Set {\rm III}
was acquired from an array of three cross-wires, arranged {\em
above} each other at heights 11 cm, 27 cm and 54 cm respectively.
These measurements are thus much closer to the desert floor.
\begin{figure}
\centerline{\includegraphics[scale=0.4]{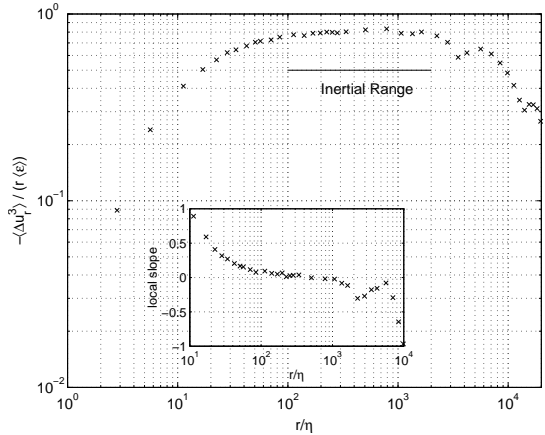}}
\caption{\label{irange}A typical compensated longitudinal
third-order structure function. The inset indicates the region of
isotropic inertial range scaling. (From Ref.~\cite{dhruva}.)}
\end{figure}
\begin{figure}
\centerline{\includegraphics[scale=0.4]{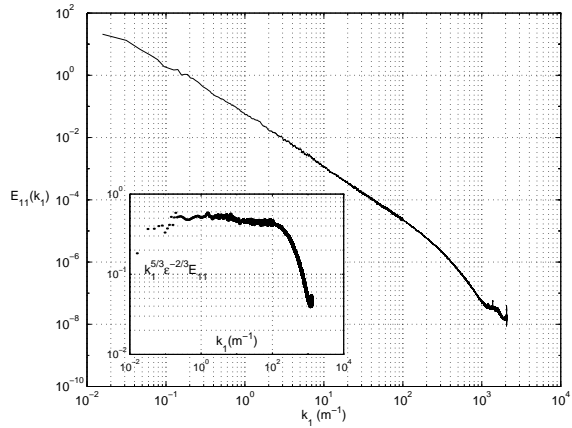}}
\caption{\label{samp_spect}The one-dimensional energy spectrum
computed from data set II. (From Ref.~\cite{dhruva}.)}
\end{figure}
The hot-wires, about 0.7 mm in length and 6 $\mu$m in diameter,
were calibrated just prior to mounting them on the meteorology
towers and checked immediately after dismounting. The hot-wires
were operated on DISA 55M01 constant-temperature anemometers.  The
frequency response of the hot-wires was typically good up to 20
kHz. The voltages from the anemometers were suitably low-pass
filtered and digitized. The voltages were constantly monitored on
an oscilloscope to ensure that they did not exceed the digitizer
limits. Also monitored on-line were power spectra from an HP 3561A
Dynamic Signal Analyzer. The wind speed and direction were
independently monitored by a direction indicator mounted on the
tower (sets I and III), or a vane anemometer a few meters away
(set II). The real-time durations of data records were limited
only by the degree of constancy demanded of the wind speed and its
direction.

The Taylor microscale Reynolds number was 10,000 for set {\rm I}
and about 20,000 for set {\rm II} \cite{dhruva}. For set {\rm
III}, the corresponding numbers were 900, 1400 and 2100,
respectively, at the three heights. Table~\ref{exp_data} lists a
few relevant facts about the data records analyzed here.
Figure~\ref{irange} shows a compensated third order longitudinal
structure function and indicates the region over which the
inertial range is expected to hold. As another example of the
nature of the data, we show in Fig.~\ref{samp_spect} the
longitudinal energy spectrum which displays extensive range of
scaling. The slope is slightly larger than 5/3, as the compensated
spectrum in the inset is supposed to clarify.
\section{Anisotropic contribution in the case of homogeneity}
\subsection{General remarks on the data}
We analyze data sets I and II described in section 3 (see Table
\ref{exp_data}) to extract the lowest order anisotropic
contribution to scaling (\cite{98ADKLPS},\cite{KLPS00}). First,
some preliminary tests and corrections need to be made. To test
whether the separation between the two probes is indeed orthogonal
to the mean wind, we computed the cross-correlation function
$\langle u_1(t+\tau)u_2(t)\rangle$.  Here, $u_1$ and $u_2$ refer
to velocity fluctuations in the direction of the mean wind, for
the arbitrarily numbered probes 1 and 2 respectively (see
Fig.~\ref{expsetup}). If the separation were precisely orthogonal
to the mean wind, the cross-correlation function should be maximum
for $\tau=0$. Instead, for set I, we found the maximum shifted
slightly to $\tau=0.022$ s, implying that the separation was not
precisely orthogonal to the mean wind. To correct for this effect,
the data from the second probe were time-shifted by 0.022 s.  This
amounts to a change in the actual value of the orthogonal
distance.  We computed this effective distance to be $\Delta
\approx 54$ cm (instead of the 55 cm that was set physically). For
set II, the effective separation distance was estimated to be 31
cm (instead of the physically set 40 cm). Next, we tested the
isotropy of the flow for separations of the order of $\Delta$.
Define the ``transverse" structure function across $\Delta$ as
$S_T(\Delta)\equiv\langle [u_1(\bar U t)-u_2(\bar Ut)]^2\rangle$
and the ``longitudinal" structure function as $S_L(\Delta)\equiv
\langle[u_1(\bar U t+\bar U t_\Delta)-u_1(\bar U t)]^2\rangle$
where $t_\Delta=\Delta/\bar U$. If the flow were isotropic we
would expect \cite{MY}
\begin{equation}
S_T(\Delta)=S_L(\Delta)+{\Delta\over 2}{\partial S_L(\Delta)\over
\partial
\Delta} \ .
\label{Stl}
\end{equation}
In the isotropic state both longitudinal and transverse components
scale with the same exponent, $S_{T,L}(\Delta)\propto
\Delta^{\zeta_2}$, and the ratio $S_T/S_L$ is computed from
(\ref{Stl}) to be $1+\zeta_2/2\approx 1.35$, because
$\zeta_2\approx 0.69$ (see below). The experimental ratio was
found to be 1.32 for set I, indicating that the anisotropy at the
scale $\Delta$ is small. This same ratio was about 1.8 for set
III, indicating higher degree of anisotropy in that scale range.
The differences between the two data sets seem attributable partly
to differences in the terrain and other atmospheric conditions,
and partly to the different distances from the ground.

Lastly, we needed to assess the effects of high turbulence level
on Taylor's hypothesis. A comparison was made of the structure
functions of two signals with turbulence levels differing by a
factor of 2 and no difference was found. The correction scheme
proposed in \cite{gustavo} also showed no changes. For a few
separation distances, the statistics of velocity increments from
two probes separated along ${\bf n}$, the direction of the mean
wind, agreed with Taylor's hypothesis. More details can be found
in Ref.~\cite{dhruva}.
\begin{figure}
\centerline{\includegraphics[scale=0.4]{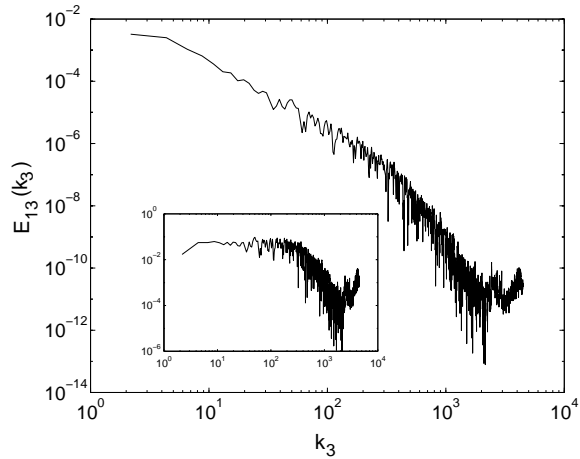}}
\caption{Log-log plot of the shear-stress cospectrum $E_{13}(k_3)$
computed at 0.54 m. The inset shows a log-log plot of
$k_3^{2.1}E_{13}(k_3)$ vs. $k_3$. The flat region indicates a
region of scaling with exponent -2.1.} \label{cospect_54}
\end{figure}

The use of the isotropy equation (\ref{Stl}) presents some
evidence regarding the prevalence of modest amounts of anisotropy
in the second-order statistics. This persistence is more explicit
if one considers another familiar object, namely the
cross-spectral density between horizontal and vertical velocity
components. We consider the one-dimensional cross-spectral density
(or shear-stress cospectrum) $E_{13}(k_3)$, which is zero in the
case of isotropy. From dimensional considerations, the scaling
exponent for this quantity is $-7/3$ (see \cite{lumley} for theory
and \cite{sv} for experimental tests). Figure~\ref{cospect_54}
shows the cospectrum computed for the height of 0.54 m. The inset
shows that the cospectrum compensated with a scaling exponent of
$-2.1$ is flat. To the extent that this is numerically smaller
than 7/3, the decay of anisotropy is slower than expected for
second-order quantities. Even allowing for the fact that the
dimensional analysis assumes K41 scaling and therefore does not
account for possible intermittency corrections in the anisotropic
sectors, it is still clear that the small scales do not attain
isotropy as fast as dimensional considerations suggest. We would
like to use the properties of the SO(3) decomposition in order to
disentangle the isotropic from the anisotropic contributions and
better quantify the anisotropy. We first consider the second-order
structure function in detail.

\subsection{The tensor form for the second-order structure function}
\subsubsection{The anisotropic tensor component derived under the assumption of axisymmetry}
To obtain a theoretical form of the structure function tensor we
first select a natural coordinate system. Our choice is to have
the mean-wind direction ${\bf n}$ along the z-axis which we label
as 3. The angle with respect to the 3-axis is the polar angle
$\theta$. The second axis is given by the separation vector ${\bf
\Delta}$ between the two probes. We make the simplifying
assumption that the main symmetry broken in the flow is the
cylindrical symmetry about the mean-wind direction. In other
words, the main anisotropic contribution is cylindrically
symmetric about the mean-wind direction. It is shown {\em a
posteriori} that this assumption probably accounts for most of the
anisotropy {\em for this particular geometrical set-up}. In the
next section, we provide a complete analysis with no assumptions
about the symmetries of the flow. We conclude from that analysis
that the simplification of the present section is not inconsistent
with the properties of the full tensor.

Next, we write down the tensor form for the general second-order
structure function (defined by Eq.~(\ref{Sn}) for $n=2$) in terms
of irreducible representations of the SO(3) rotation group. Since
we are far from the wall, our interest is in relatively modest
anisotropies, and so we focus on the lowest order correction to
the isotropic ($j=0$) contribution. We therefore write
\begin{equation}
S^{\alpha\beta}({\bf r}) = S_{j=0}^{\alpha\beta}({\bf r}) +
S_{j=2}^ {\alpha\beta}({\bf r})+\dots
\end{equation}
We do not have a $j=1$ term because it vanishes from parity
considerations (the structure function itself is even in $r$ in
our experimental configuration) or by the incompressibility
constraint. Assumed cylindrical symmetry of the anisotropic $j=2$
contribution implies that we only consider the $m=0$ subspace in
this contribution. Now, the most general form of the tensor can be
written down by inspection. The case $j=0$ is well known (see
Eq.~(\ref{S2_incomp})) to be
\begin{equation}
S_{j=0}^{\alpha\beta}({\bf r})=c_0(r) \left[(2+\zeta_2) \delta^
{\alpha\beta}-\zeta_2{r^\alpha r^\beta\over r^2}\right]\,
\end{equation}
where $c_0(r)=c_0r^{\zeta_2}$, and $c_0$ is a non-universal
numerical coefficient that needs to be obtained from fits to the
data. The $j=2$ component has six independent tensor forms and
corresponding coefficients. These correspond to the different
$a_{q_{00}}$ described in section 1. As with $j=0$, the
$j=2$ component can be simplified by imposing conditions of
incompressibility and orthogonality with the $j=0$ part of the
tensor. This leaves us (in the case of cylindrical symmetry) with
two independent coefficients which we call $a$ and $b$, giving,
\begin{eqnarray}
S^{\alpha\beta}_{j=2} ({\bf r}) &=&
ar^{\zeta_2^{(2)}}\Big[(\zeta_2^{(2)}
-2)\delta^{\alpha\beta} -
\zeta_2^{(2)}(\zeta_2^{(2)}+6)\nonumber\\ &\times&\delta^{\alpha\beta}
{({\bf n}\cdot {\bf r})^2 \over r^2}+2\zeta_2^{(2)}(\zeta_2^{(2)}-2){r^\alpha
r^\beta({\bf n}\cdot {\bf r})^2
\over r^4}\nonumber\\
&+&([\zeta_2^{(2)}]^2+3\zeta_2^{(2)}+6)n^\alpha n^\beta \nonumber\\
&-&{\zeta_2^{(2)}(\zeta_2^{(2)}-2)\over r^2}(r^\alpha n^\beta +
r^\beta n^\alpha)(\bf n\cdot \bf r)\Big]\label{finalform}\\
&+& br^{\zeta_2^{(2)}}\Big
[-(\zeta_2^{(2)} +3)(\zeta_2^{(2)}+2)\delta^ {\alpha\beta}(\bf n\cdot
\bf r)^2\nonumber\\
&+&(\zeta_2^{(2)} - 2){r^\alpha r^\beta \over r^2} +(\zeta_2^{(2)} +3)
(\zeta_2^{(2)}+2)n^\alpha n^\beta\nonumber\\ &+& (2\zeta_2^
{(2)}+1)(\zeta_2^{(2)}-2)
{{r^\alpha}{r^\beta}{({\bf n}\cdot{\bf r})^2} \over r^4}\nonumber\\
&-&([\zeta_2^{(2)}]^2 - 4)(r^\alpha n^\beta + r^\beta
n^\alpha)(\bf n\cdot \bf r)\Big]. \nonumber
\end{eqnarray}
We note that the forms for the tensor were derived on the basis of
the assumption of cylindrical symmetry. We used neither the
Clebsch-Gordon method nor the ``adding indices'' method of section
1. This is why they were not automatically orthogonal to the $j=0$
contribution; that condition had to be explicitly imposed.
However, in the end, this method also yields the same
number of independent coefficients as the formal Clebsch-Gordon
methods. This merely shows that there are different ways of
writing the tensor forms for the basis elements. However, the
Clebsch-Gordon rules tell us the number, parity and symmetry of
the terms we must in the end have for a given $j,m$ sector.

Finally, we note that in the present experimental set-up only the
component of the velocity in the direction of ${\bf n}$ (the
3-axis) is measured. In the coordinate system chosen above we can
read from Eqs. (\ref{S2_incomp}) and (\ref{finalform}) the
relevant component to be
\begin{eqnarray}
S^{33}(r,\theta)&=&S^{33}_{j=0}(r,\theta)+ S^{33}_{j=2}(r,\theta)
\nonumber\\
&=&c_0\left({r\over \Delta}\right)^{\zeta_2} \Big[ 2+\zeta_2-\zeta_2
\cos^2\theta\Big]\nonumber\\
&+&a\left({r\over \Delta}\right)^{\zeta_2^{(2)}}
\Big[ (\zeta_2^{(2)}+2)^2-\zeta_2^{(2)}
(3\zeta_2^{(2)}+2)\cos^2\theta\nonumber\\ &+& 2\zeta_2^{(2)}(\zeta_2^{(2)}
-2)\cos^4\theta\Big] \label{ourcomp}\\
&+&b\left({r\over
\Delta}\right)^{\zeta_2^{(2)}}\Big[(\zeta_2^{(2)}+2) (\zeta_2^{(2)}+3)-
\zeta_2^{(2)}(3\zeta_2^{(2)}+4)\nonumber\\
&\times&\cos^2\theta+(2\zeta_2^{(2)}+1)
(\zeta_2^{(2)}-2)\cos^4\theta\Big] \ . \nonumber
\end{eqnarray}
Here $\theta$ is the angle between ${\bf r}$ and ${\bf n}$, and $r$ has
been normalized by $\Delta$, making all the coefficients dimensional,
with units of (m/sec)$^2$. Taylor's hypothesis allows us to obtain
components of $S^{\alpha \beta}$ from
(\ref{ourcomp}), with $\theta=0$ and
variable, respectively. In other words,
\begin{equation}
S^{33}(r,\theta=0)=\langle [u_1
(\bar U t+\bar Ut_r)-u_1(\bar Ut)]^2\rangle\ ,
\label{S33_theta0}
\end{equation}
where $t_r\equiv r/\bar U$, and
\begin{equation}
S^{33}(r,\theta)=\langle [u_1(\bar U t +\bar Ut_{\tilde
r})-u_2(\bar Ut)]^2\rangle.
\label{S33_theta}
\end{equation}
Here
$\theta=\arctan(\Delta/\bar Ut_{\tilde r})$, $t_{\tilde r}=\tilde
r/\bar U$, and $r=\sqrt{\Delta^2+(\bar Ut_{\tilde r})^2}$.

\begin{figure}
\centerline{\includegraphics[scale=0.9]{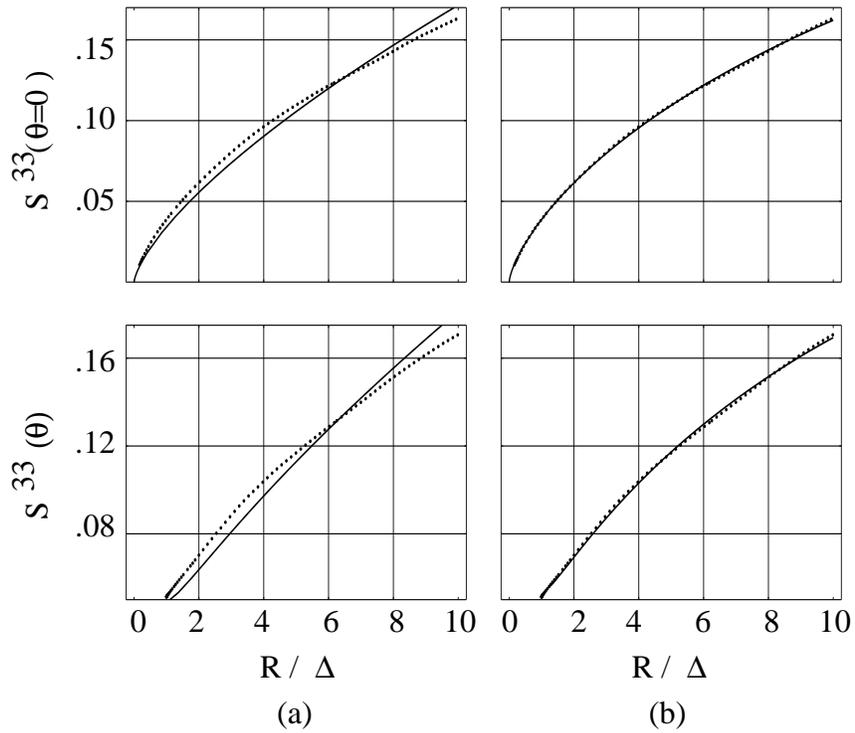}}
\caption{The structure functions $S^{33}$ for $\theta=0$ and for
non-zero $\theta$ computed for set I. The dots are for
experimental data and the line is the analytic fit. Panel (a)
presents fits to the $j=0$ component only, and panel (b) to
components $j=0$ and $j=2$ together.} \label{Fig.1}
\end{figure}

\begin{figure}
\centerline{\includegraphics[scale=0.9]{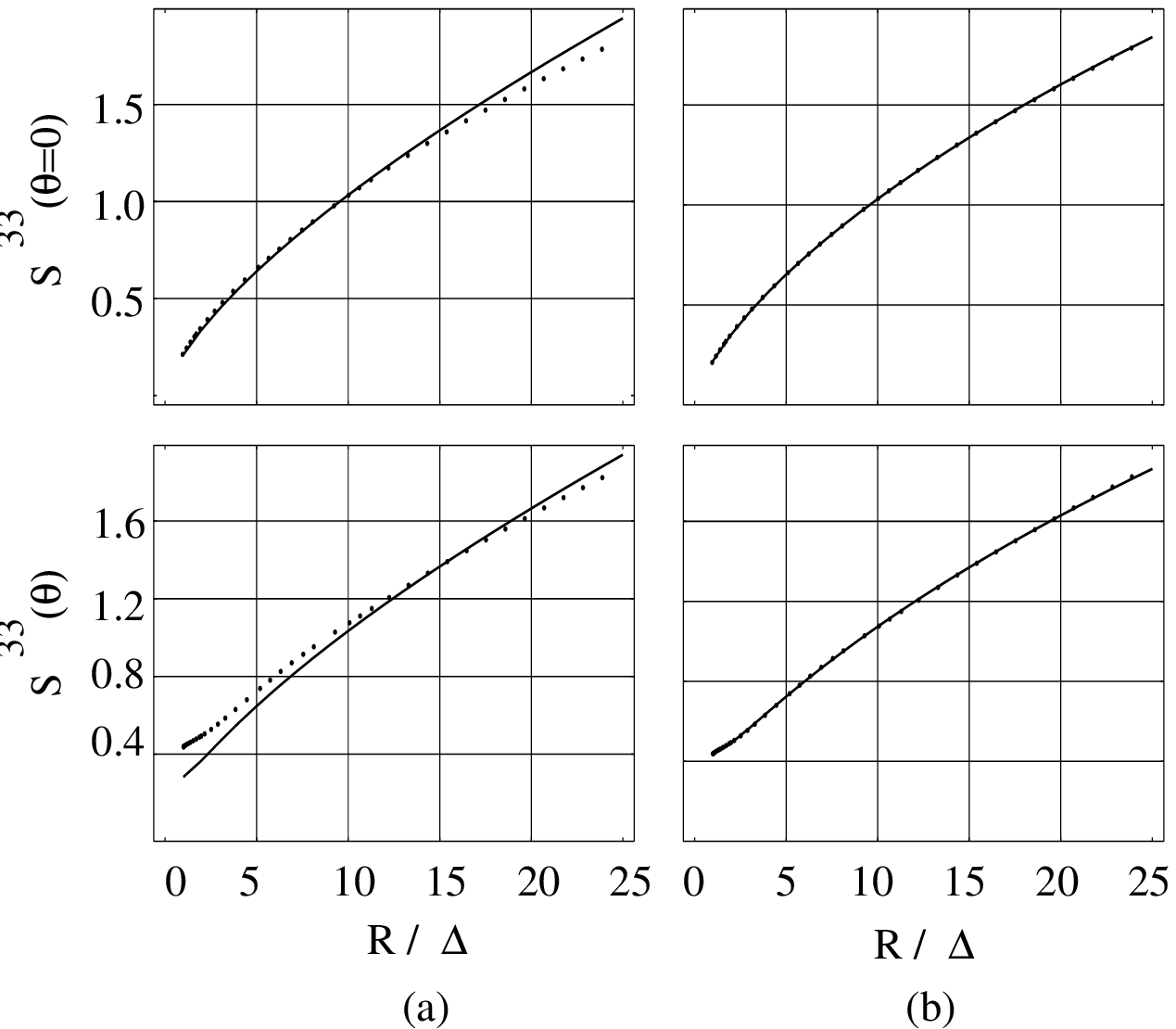}}
\caption{The structure functions $S^{33}$ for $\theta=0$ and for
non-zero $\theta$ computed for set III. The dots are for
experimental data and the line is the analytic fit. Panel (a)
presents fits to the $j=0$ component only, and panel (b) to the
sum of the components $j=0$ and $j=2$.} \label{Fig.2}
\end{figure}

The quantities on the left hand side of Eqs.~(\ref{S33_theta0})
and (\ref{S33_theta}) were computed from the experimental data and
fitted to the theoretical expression (\ref{ourcomp}) using the
appropriate values of $\theta$.  The fits were performed in the
range $1 < r/\Delta < 10$ ($0.54$ m $< r < 5.4$ m) for set I and
$1 < r/\Delta < 25$ ($0.31$ m $< r < 8$ m) for set II. The ranges
were based on the constancy of the third order structure function
(see Fig.~\ref{irange} for an example). Panels (a) of
Figs.~\ref{Fig.1} and \ref{Fig.2} show, for data sets I and II
respectively, a comparison between the measured $S^{33}(r,
\theta=0)$ and the $j=0$ form of the equation.  The comparison
shows that the agreement is modest, and that the best-fit yields
the exponent $\zeta_2$ to be $0.69$. A careful analysis of the
data elsewhere \cite{sreeni} shows that, if the effect of the
shear is removed in a plausible way, the power-law fit is
excellent over a range of scale separations. To include the $j=2$
contribution, we fixed $\zeta_2$ to be 0.69 and performed the
following analysis. For given values of the variables $r$ and
$\theta$, we guessed the second exponent $\zeta_2^{(2)}$ and
estimated the unknown coefficients $c_0$, $a$ and $b$ by using a
linear regression algorithm.  We followed this procedure
repeatedly for different values of $\zeta_2^{(2)}$ ranging from
$0$ to $2$. We then chose the value of $\zeta_2^{(2)}$ that
minimized $\chi^2$ (the sum of the squares of the differences
between the experimental data and the fitted values). 

\begin{figure}
\centerline{\includegraphics[scale=0.6]{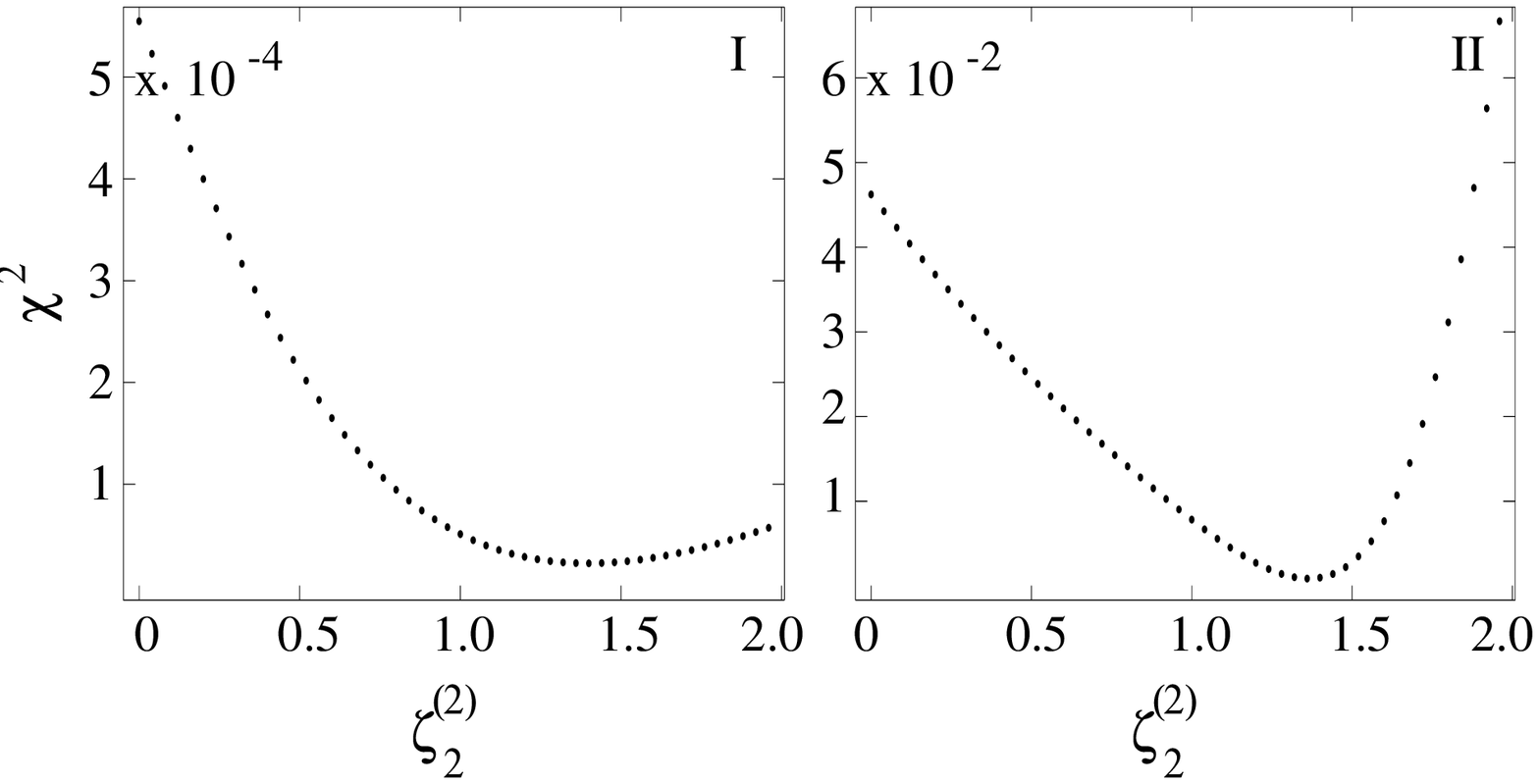}}
\caption{\label{Fig.3}The determination of the exponent
$\zeta_2^{(2)}$ from a least-square fit of $S^{33}(r,\theta)$ to
its analytic form. From set I we obtain a numerical value of the
best fit, $\zeta_2^{(2)}=1.38\pm 0.15$ while from set III the best
fit value is $1.36\pm 0.1$ both of which are in close agreement
with the theoretical expectation of 4/3 of section 2.4 (without
intermittency corrections). The differences in the nature of the
minima are not understood.}
\end{figure}

In Fig.~\ref{Fig.3} we present $\chi^2$ values as a function of
$\zeta_2^{(2)}$. The optimal value of this exponent and the
uncertainty determined from these plots is $\zeta_2^{(2)}\approx
1.38\pm 0.15$ from set I, and $\zeta_2^{(2)}\approx 1.36\pm 0.1$
from set II. The best numerical values for the coefficients are
presented in Table~\ref{j2fit_results}. Panels (b) in Figs.
\ref{Fig.1} and \ref{Fig.2} show fits to the sum of the $j=0$
and $j=2$ contributions to the experimental data. Even though the
$j=2$ contributions are small, they improve the fits tremendously.
This situation lends support to the essential correctness of the
present analysis.
\begin{table}
\centerline{\begin{tabular} {|c|c|c|c|c|}\hline
$\zeta_2$&$\zeta_2^{(2)}$&$c_0$&$a$&$b$\\ \hline
0.69&1.38$\pm0.15$&0.023$\pm0.001$&-0.0051$\pm0.0006$&0.0033$\pm0.0005$\\
\hline
0.69&1.36$\pm0.10$&0.112$\pm0.001$&-0.052$\pm0.004$&0.050$\pm 0.004$\\ \hline
\end{tabular}}
\caption{The scaling exponents and the three coefficients in units
of (m/sec)$^2$ as determined from the nonlinear fit of Eq.~(7) to
data sets I (first line) and II (second line).}
\label{j2fit_results}
\end{table}
The figures show that the purely longitudinal structure function
corresponding to $\theta=0$ is somewhat less affected by the
anisotropy than is the finite $\theta$ structure function (see
especially, Fig.~\ref{Fig.2}). The reason is the closeness of the
numerical absolute values of the coefficients $a$ and $b$ (see
Table \ref{j2fit_results}). For $\theta=0$ the two tensor
forms multiplied by $a$ and $b$ coincide, and the $j=2$
contribution becomes very small. The value of $\zeta_2 = 0.69$
quoted above can be obtained from such a fit to the $\zeta_2 = 0$
part alone; as long as one measures only this component, it seems
reasonable to proceed with just that one exponent. However, the
inclusion of the second exponent $\zeta_2^{(2)}$ improves the fit
even for the longitudinal case; for the finite $\theta$ case, this
inclusion appears essential for a good fit. In fact, the fit using
$j=0$ and $j=2$ for data set II extends all the way to 40 m
(beyond the range shown in Fig.~\ref{Fig.2}). This indicates that
the ``inertial-range'' where scaling theory applies is much longer
than anticipated by traditional log-log plots.  The close
agreement with the theoretical expectation of 4/3 (e.g.,
Refs.~\cite{lumley},\cite{KLPS00}), and the apparent
reproducibility of the result for two different experiments is a
strong indication that this exponent may be universal.

It should be understood that, for the objects considered in this 
section, the exponent $\zeta_2^{(2)}$ is just the smallest exponents 
in a possible
hierarchy $\zeta_2^{(j)}$ that characterizes higher order
irreducible representations indexed by $j$. (The case of $\zeta_2^{(1)}$
will be discussed in section 5.) As discussed earlier,
we expect the exponents to be a non-decreasing function of $j$,
and that the highest values of $j$ are being peeled off quickly
when $r$ decreases. Nevertheless, the lower order values of
$\zeta_2^{(j)}$ can be measured and computed. The results of this
section were first reported in \cite{98ADKLPS}. Results are also
in agreement with the subsequent analysis of numerical simulations
\cite{99ABMP}.

\subsubsection{The complete $j=2$ anisotropic contribution}
In the previous section, preliminary results on the scaling
exponent $\zeta_2^({2})$, were obtained under the assumption of
cylindrical symmetry of the dominating anisotropic contribution.
The analysis here is more complete, and takes into account the
full tensorial structure. We show that this is feasible, and the
final results are in agreement with those presented in the
previous section.

The method used to extract the unknown anisotropic scaling
exponent is essentially the same as in the previous section. Since
we look for the lowest order anisotropic contributions in our
analyses, we perform a two-stage procedure to separate the various
sectors. First we look at the small scale region of the inertial
range to determine the extent of the fit with a single (isotropic)
exponent. We then seek to extend this range by including
appropriate anisotropic tensor contributions, and obtain the
additional scaling exponents using least-squares fitting
procedure. This procedure is self-consistent.

Reference \cite{99LPP} presents a detailed analysis of the
consequences of Taylor's hypothesis on the basis of an exactly
soluble model. It also proposes ways for minimizing the systematic
errors introduced by the use of Taylor's hypothesis. In light of
that analysis we will use an ``effective" wind, $U_{\rm eff}$, for
surrogating the time data. This velocity is a combination of the
mean wind $\overline{U}$ and the root-mean-square $u'$,
\begin{equation} U_{\rm eff}
\equiv \sqrt{\overline{U}^2+(bu')^2} \, \label{defUeff}
\end{equation}
where $b$ is a dimensionless parameter.

In the second order structure function defined already, {\em viz.},
\begin{equation}\label{Sab}
S^{\alpha\beta}({\bf r}) = \langle (u^\alpha({\bf x} + {\bf r}) -
u^\alpha({\bf x})) (u^\beta({\bf x} + {\bf r}) - u^\beta({\bf
x}))\rangle , \end{equation} the $j=2$ component of the SO(3)
symmetry group corresponds to the lowest order anisotropic
contribution that is symmetric in the indices, and has even parity
in ${\bf r}$ (due to homogeneity). Although the assumption of
axisymmetry used in \cite{98ADKLPS} and in the previous section
seemed to be justified from the excellent qualities of fits
obtained, we attempt to fit the same data (set I) with the {\em
full} tensor form for the $j=2$ contribution. The derivation of
the full $j=2$ contribution to the symmetric, even parity,
structure function appears in appendix A. We then find the range
of scales over which the structure function
\begin{equation}\label{tta0}
S^{33}(r,\theta = 0) = \langle (u_1^{(3)}(x+r) - u_1^{(3)}(x))^2
\rangle,
\end{equation}
with the subscript $1$ denoting one of the two probes, can be
fitted with a single exponent. To find the $j=2$ anisotropic
exponent we need to use data taken from both probes. To clarify
the procedure, for the geometry shown in Fig.~\ref{expsetup}, what
is computed is actually
\begin{equation}
S^{33}(r,\theta)=\langle [u^{(3)}_1( U_{\rm eff} t + U_{\rm
eff}t_{\tilde r})- u^{(3)}_2( U_{\rm eff}t)]^2\rangle.
\end{equation}
Here $\theta=\arctan(\Delta/ U_{\rm eff}t_{\tilde r})$, $t_{\tilde
r}=\tilde r/ U_{\rm eff}$, and $r=\sqrt{\Delta^2+(\bar U_{\rm
eff}t_{\tilde r})^2}$. $U_{\rm eff}$ is defined by
Eq.~(\ref{defUeff}) with the optimal value of $b$ taken from model
studies to be 3. For simplicity we shall refer from now on to such
quantities as
\begin{equation}\label{ttadep}
S^{33}(r,\theta) = \langle (u_1^{(3)}(x+r) - u_2^{(3)}(x))^2\rangle \ .
\end{equation}

\begin{figure}
\centerline{\includegraphics[scale=0.61]{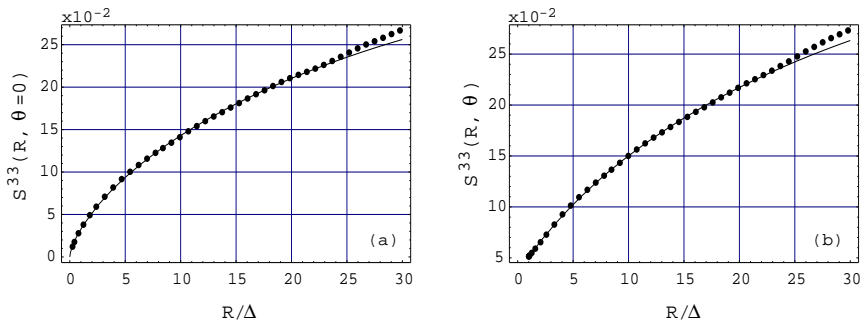}}
\caption{The structure functions computed from data set I and fit with the
$j=0$ and full $j=2$ tensor contributions using the best fit values
of exponents $\zeta_2=0.68$ and $\zeta_2^{(2)}=1.38$.
Panel (a) shows the fit to
the single-probe ($\theta = 0$) structure function in the range
$0.2 < r/\Delta < 25$ and panel (b) shows the fit to the
$\theta$-dependent structure function in the range $1 < r/\Delta < 25$.}
\label{sfj2full}
\end{figure}

Next, we fix the scaling exponent of the isotropic sector to be $0.69$
and find the $j=2$ anisotropic exponent that results from fitting to
the full $j=2$ tensor contribution. We fit the objects in
Eqs.~(\ref{tta0}) and (\ref{ttadep}) to the sum of the $j=0$ (with
scaling exponent $\zeta_2 = 0.69$) and the $j=2$ contributions. The sum 
is given by (see appendix A)
\begin{eqnarray}
S^{33}(r,\theta)&=&S^{33}_{j=0}(r,\theta)+ S^{33}_{j=2}(r,\theta) \nonumber\\
&=&c_0\left({r\over \Delta}\right)^{\zeta_2} \Big[ 2
+\zeta_2-\zeta_2 \cos^2\theta\Big] \nonumber\\
&+&a\left({r\over \Delta}\right)^{\zeta_2^{(2)}}
\Big[ (\zeta_2^{(2)}+2)^2 -\zeta_2^{(2)}
(3\zeta_2^{(2)}+2)\cos^2\theta\nonumber\\
&&\mbox{~}\hspace{1.7cm}+2\zeta_2^{(2)}(\zeta_2^{(2)}-2)\cos^4\theta\Big] \nonumber\\
&+&b\left({r\over
\Delta}\right)^{\zeta_2^{(2)}} \Big[ (\zeta_2^{(2)}+2) (\zeta_2^{(2)}+3)-
\zeta_2^{(2)}(3\zeta_2^{(2)}+4)\cos^2\theta\nonumber\\
&&\mbox{~}\hspace{1.7cm}+(2\zeta_2^{( 2)}+1) (\zeta_2^{(2)}-2)\cos^4\theta\Big] \\
&+&a_{9,2,1} \left({r\over \Delta}\right)^{\zeta_2^{(2)}}
\Big[-2\zeta_2^{(2)} (\zeta_2^{(2)}+2)
\sin\theta\cos\theta \nonumber\\
&&\mbox{~}\hspace{2.35cm}+ 2\zeta_2^{(2)}(\zeta_2^{(2)}-2)\cos^3\theta\sin\theta \Big]\nonumber\\
&+&a_{9,2,2} \left({r\over
\Delta}\right)^{\zeta_2^{(2)}}\Big[-2\zeta_2^{(2)} (\zeta_2^{(2)}-2)
\cos^2\theta\sin^2\theta\Big]
\nonumber\\
&+&a_{1,2,2}\left({r\over \Delta}\right)^{\zeta_2^{(2)}}
\Big[-2\zeta_2^{(2)} (\zeta_2^{(2)}-2)\sin^2\theta\Big].\nonumber
\label{fulltens}
\end{eqnarray}
We fit the experimentally generated functions to the above form
using values of $\zeta_2^{(2)}$ ranging from $0.5$ to $3$.  Each
iteration of the fitting procedure involves solving for the six
unknown, non-universal coefficients. The best value of
$\zeta_2^{(2)}$ is the one that minimizes the $\chi^2$ value for
these fits; we obtain that to be $1.38\pm0.15$. The fits with this
choice of exponent are displayed in Fig.~\ref{sfj2full}.

The corresponding values of the six fitted coefficients is given
in Table~2. The range of scales that are fitted to this
expression is $0.2 < r/\Delta < 25$ for the $\theta=0$
(single-probe) structure function and $1 < r/\Delta < 25$ for the
$\theta \ne 0$ (two-probe) structure function.  We are unable to
fit Eq.~(\ref{fulltens}) to scales larger
than about 12 meters without losing the quality of the fit in
the small scales. This limit is roughly twice the height of the probe
from the ground. Based on the earlier discussion, we should be in
the regime of the largest scales up to which the three-dimensional
theory would hold. (Beyond this scale the boundary layer is
dominated by the sloshing motions which are quasi two-dimensional
in nature.) Therefore this limit to the fitting range is
consistent with our expectations for the maximum scale of
three-dimensional turbulence. We conclude that the structure
functions exhibit scaling behavior over the whole scaling range,
but this important fact is missed if one does not consider a
superposition of the $j=0$ and $j=2$ contributions. We thus
conclude that the estimate for the $j=2$ scaling exponent
$\zeta_2^{(2)}\approx 1.38$. This same estimate was obtained in
\cite{98ADKLPS} and in the previous section using only the
axisymmetric terms. The value of the coefficients $a$ and $b$ are
again close in magnitude but opposite in sign---just as in
\cite{98ADKLPS}, giving a small contribution to
$S^{33}(r,\theta=0)$. The non-axisymmetric contributions vanish in
the case of $\theta = 0$. The contribution of these terms to the
finite $\theta$ function is relatively small because the angular
dependence appears as $\sin\theta$ and $\sin^2\theta$, both of
which are small for small $\theta$ (large $r$); it was for this
reason that we were able to obtain in the previous section a good
fit to just the axisymmetric contribution. Lastly, we note that
the total number of free parameters in this fit is 7 (6
coefficients and 1 exponent). The relative ``flatness'' of the
$\chi^2$ function near its minimum (see, especially the left panel
of Fig.~\ref{Fig.3}) may be indicative of the large number of free
parameters in the fit. However, the value of the exponent is
perfectly in agreement with the analysis of numerical simulations
\cite{99ABMP} in which one can properly integrate the structure
function against the basis functions, eliminating all
contributions except that of the $j=2$ sector. Furthermore, fits
to the data in the vicinity of $\zeta_2^{(2)} = 1.38$ show enough
divergence from experiment that we are satisfied about the
genuineness of the $\chi^2$ result. The results of this analysis
taking into account the full $j=2$ tensor were first presented in
\cite{KLPS00}.
\subsection{Summary}
We consider the lowest order anisotropic contribution to the
second order tensor functions of velocity in the atmospheric
boundary layer. The $j=1$ contribution is absent in our
experimental configuration because of the incompressibility
constraint. The $j=2$ sector is expected to be the dominant
contribution to anisotropy. First we make the {\it a priori}
assumption that the cylindrical symmetry is broken in the first
deviation from anisotropy. The use of the Clebsch-Gordon rules
tell us the number of terms that must be included. We derive on this
basis the tensor form for the $j=2$ contribution that is
axisymmetric. The conditions of orthogonality with $j=0$ and
incompressibility are used to constrain the unknown coefficients
giving two unknown coefficients and one unknown exponent in the
anisotropic sector. Using this form along with the known isotropic
contribution, we can extract the anisotropic scaling exponent from
the experimental data. We have also used the full form of the
$j=2$ tensor including all its $2j+1$ components in order to
examine if the results from the initial analysis were justified.
The excellent agreement between the two strengthens our confidence
in the value of the $j=2$ exponent $\zeta^{(2)}_2 \approx 1.36$,
and in our initial assumption of the breaking of 
axisymmetry of the flow.
\section{Anisotropic contribution in the case of inhomogeneity}
\subsection{Extracting the j=1 component}
The homogeneous structure function defined in Eq.~(\ref{Sab})
is known from properties of
symmetry and parity to possess no contribution from the $j=1$ sector (see
section B.2), the $j=2$ sector being its lowest order
anisotropic contributor. In order to isolate the scaling behavior of
the $j=1$ contribution in atmospheric shear flows we must either
($a$) explicitly construct a new tensor object which will allow for such
a contribution, or ($b$) extract it from  the structure function
itself computed in the presence of {\em inhomogeneity}.
Adopting the former approach, we construct the tensor
\begin{equation}\label{Tab}
T^{\alpha\beta}({\bf  r}) = \langle u^\alpha({\bf  x} + {\bf  r}) -
u^\alpha({\bf  x}))(u^\beta({\bf  x} + {\bf  r}) + u^\beta({\bf  x}))\rangle.
\end{equation}
This object vanishes both when $\alpha=\beta$ and when ${\bf  r}$
is in the direction of homogeneity, viz., the streamwise
direction. From data set III we can calculate this function for
non-homogeneous scale-separations (in the shear direction). In
general, this will exhibit mixed parity and symmetry; the incompressibility 
condition does not reduce our parameter space.
Therefore, to minimize the final number of fitting parameters, we
examine only the antisymmetric contribution. In section B.2, we
have derived the antisymmetric tensor contributions in the $j=1$
sector, and used this to fit for the unknown $j=1$ exponent. We
describe the results of this effort below. This can be used to find
$j=1$ exponent for the inhomogeneous structure function which is
symmetric but has mixed parity. We do not present the result of
that analysis here essentially because it is consistent with those
from the antisymmetric case.

Returning to consideration of the antisymmetric part of the tensor
object defined in Eq.~(\ref{Tab}), {\em viz.},
\begin{eqnarray}
{\widetilde T}^{\alpha\beta}({\bf  r}) &=&
{T^{\alpha\beta}({\bf  r}) - T^{\beta\alpha}({\bf  r}) \over 2}
\nonumber\\
&=&\langle u^\alpha({\bf  x})u^\beta({\bf  x} + {\bf  r})\rangle
- \langle u^\beta({\bf  x})u^\alpha({\bf  x} + {\bf  r})\rangle,
\end{eqnarray}
it is easy to see that it will only have contributions from the
antisymmetric $j=1$ basis tensors. An additional useful property
of this object is that, for the configuration of data set II, it
does not have any contribution from the isotropic $j=0$ sector
spanned by $\delta^{\alpha\beta}$ and $r^\alpha r^\beta$ since
these objects are symmetric in the indices. This allows us to
isolate the $j=1$ contribution and determine its scaling exponent
$\zeta_2^{(1)}$ starting from the smallest scales available. Using
data (set III) from the probes at $0.27$~m (probe 1) and at
$0.11$~m (probe 2) we calculate
\begin{equation}
{\widetilde T}^{31}({\bf  r}) = \langle u_2^{(3)}({\bf
x})u_1^{(1)}({\bf x} + {\bf r})\rangle - \langle u_1^{(3)}({\bf x}
+ {\bf r})u_2^{(1)}({\bf x})\rangle\,
\end{equation}
where again superscripts denote the velocity component and
subscripts denote the probe by which this component is measured.
The goal is to fit this experimental object to the tensor form
derived in appendix B, Eq.~(\ref{T31_final}), namely,
\begin{eqnarray}\label{j1tens}
{\widetilde T}^{31}(r,\theta,\phi=0) =
- a_{3,1,0}r^{\zeta_2^{(1)}}\sin\theta
+ a_{2,1,1}r^{\zeta_2^{(1)}}
+ a_{3,1,-1}r^{\zeta_2^{(1)}}\cos\theta.
\end{eqnarray}

\begin{figure}
\centerline{\includegraphics[scale=0.65]{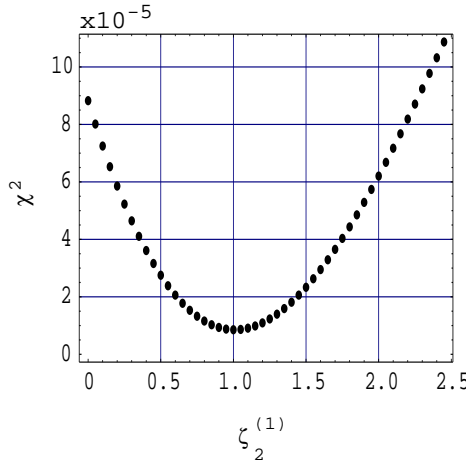}}
\caption{The $\chi^2$ minimization by the best-fit value of the exponent
$\zeta_2^{(1)}$ of the $j=1$ anisotropic sector from the fit to
$\theta$-dependent ${\widetilde T}^{31}(r,\theta)$ function in
the range $1 < r/\Delta < 2.2$. }
\label{chj1T}
\end{figure}

\begin{figure}
\centerline{\includegraphics[scale=0.75]{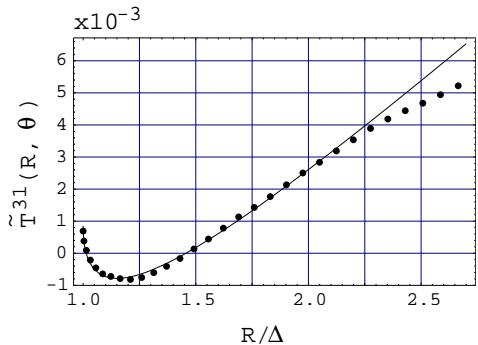}}
\caption{The fitted ${\widetilde T}^{31}(r,\theta)$ function.
The dots indicate the data and the line is the fit.}
\label{Tfit}
\end{figure}

Figure~\ref{chj1T} gives the $\chi^2$ minimization of the fit as a function of
$\zeta_2^{(1)}$. We obtain the best value to be $1 \pm 0.15$ for the final fit.
This is shown in Fig.~\ref{Tfit}.
\begin{table}
\centerline{\begin{tabular}{|c|c|c|c|}\hline
$\zeta_2^{(1)}$ & $a_{3,1,0} $ & $a_{2,1,1}$ & $a_{3,1,-1}$
\\ \hline $1 \pm 0.15 $ & $ 0.0116 \pm 0.001 $ & $ 0.0124 \pm 0.001 $ &
$-0.0062 \pm 0.001$\\\hline
\end{tabular}}
\caption{\label{j1_results}The values of the exponents and coefficients (in units of
(m/sec)$^2$)
obtained from the fit to the function ${\widetilde T}^{31}(r,\theta)$.}
\end{table}
The fit in Fig.~\ref{Tfit} peels off at around $r/\Delta = 2$. The
values of the coefficients corresponding to the exponent
$\zeta_2^{(1)}=1$ are given in Table~\ref{j1_results}. The maximum
range of scales over which the fit works is of the order of the
height of the probes from the ground, consistent with the
considerations presented earlier. This value of the scaling
exponent of the $j=1$ sector is in agreement with the
theoretically expected value of unity (section 2.4). Again, we
have satisfied ourselves that a different value of the exponent
yields a substantially poorer fit to the data. These findings
significantly strengthen the proposition \cite{99ALP} that the
scaling exponents in the various sectors (at least up to $j=2$)
are likely to be universal.
\section{The higher order structure functions}
\subsection{Introduction}
The method of SO(3) decomposition is applicable to tensor
functions of any rank. However, computing the anisotropic terms in
the higher order structure functions ($n>2$) becomes an
increasingly difficult task. The number of contributions to a
given $j$ (i.e., the $q$ index) quickly becomes too large. If we
were to use the methods adopted in the case of the second order
structure function, the large number of free parameters available
for fitting the data guarantees over-fitting. This would not allow
us to unambiguously extract the anisotropic scaling behavior.  In
order to circumvent this problem we recall two properties of the
tensor decomposition. First, for the single-point measurements,
with $\theta=0$ (no angular dependence), although the tensorial
information is lost, the free coefficients collapse to a single
number, see Eqs.~(\ref{j0_singlept}) and (\ref{j2_singlept}).
Second, the $isotropic$ part of the structure function tensor is
very easily obtained, as we will show below. Examination of this
part of the SO(3) expansion will show us which tensor components
do not contribute to the isotropic sector. We then extract
anisotropic exponents by considering only those tensor components
that are explicitly zero in the isotropic sector, so that whatever
is measured derives its contribution $entirely$ from the
anisotropic sector. We use an interpolation formula to compensate
for the large-scale encroachment of inertial-range scales. This
allows us to examine the lowest order anisotropic scaling
behavior. The resulting anisotropic exponents for a given
tensorial order are larger than those known for the corresponding
isotropic part. One conclusion that emerges is that the anisotropy
effects diminish with decreasing scale, although more slowly
than previously thought.

We can use the present method in principle to examine the
anisotropic contribution of tensors of $any$ order without
requiring the knowledge of the particular mathematical form of the
anisotropic sectors of these tensors. This is a considerable
advantage theoretically because the high-order tensors are
non-trivial to compute; it is an advantage experimentally because,
unlike in numerical simulations, one can measure only some
components for simple geometric arrangements of probes. The
results of this analysis were first presented in \cite{KS00}.

\subsection{Method and results}

In this part of the analysis we use data set II described in
section 2. We first consider the second-order tensor
$S^{\alpha\beta}({\bf r})$. Isotropy implies that this tensor can
be expressed as a linear combination of two terms,
$\delta^{\alpha\beta}$ and $ r^\alpha r^\beta$. As is well known,
both terms give non-zero contributions to longitudinal as well as
transverse components, corresponding to $\alpha = \beta$. For
$\alpha \ne \beta$ these two terms are identically zero if ${\bf
r}$ is taken to be in the streamwise direction 3. Therefore, we
compute the so-called mixed structure function
\begin{equation}
S^{31}(r) = \langle (u^3(x+r) - u^3(x))(u^1(x+r) - u^1(x))
\rangle, \label{S13}
\end{equation}
where, as already noted, the superscripts 1 and 3 denote the
vertical and streamwise components respectively. This object is
identically zero in the isotropic sector, and so, any non-zero value
comes from anisotropy. In particular, any scaling
behavior that it obeys should relate solely to anisotropy. By
computing Eq.~(\ref{S13}) and examining its scaling, we intend to
extract the purely anisotropic scaling behavior in the $j=2$
sector, uncontaminated by any isotropic scaling, in contrast to
the case of either longitudinal or transverse structure functions.

\subsubsection{The second-order structure function}
The previous paragraph provides the motivation for examining the
measured structure functions $S^{31}(r)$. However, as we shall see
shortly, apart from the expected $r^2$ behavior in the dissipative
range and saturation at some large scale, there appears to be no
distinct inertial range scaling. We suspect that this happens
because there is poor scale separation, since the probes are
fairly close to the ground; in fact, the large scales (which we
expect to be of the order of the height of the probe from the
ground and larger (see previous section)) may be encroaching
significantly into the inertial range. We would be aided
materially in our search for scaling if, somehow, the large-scale
effects can be separated. One way of doing this is to write down
an interpolation function that models the entire structure
function in its three different scaling regions---a dissipative
range that scales like $r^2$ when $r$ is of the order of the
Kolmogorov scale $\eta$, a large-scale behavior that tends to
saturate (indicating decorrelation) as $r$ gets to be larger than
$L$, and the intermediate inertial range for $\eta << r << L$
which may exhibit scaling. Through the use of the interpolation
formula, one can extract the scaling part in a natural way. This
is described below.

\begin{figure}
\centerline{\includegraphics[scale=0.4]{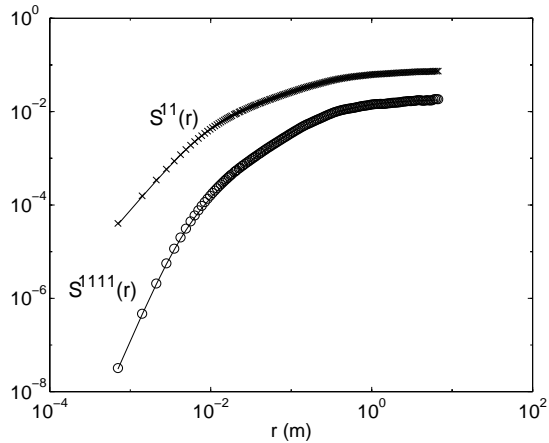}}
\caption{Log-log plots of transverse structure functions at 0.54m.
$\times$ denote the second order, O the fourth order and the solid lines
represent the interpolation fit.} \label{interp_test}
\end{figure}

A suitable form of the interpolation function is given in
\cite{dhruva} for structure functions of arbitrary order. It has
the form
\begin{equation}
S^{\alpha_1\alpha_2...\alpha_n}(r) = {A_n
\eta^n(r/\eta)^n\over(1+B_n(r/\eta)^2)^{C_n}}(1+D_n(r/L))^{2C_n-n},
\label{interp}
\end{equation}
where $A_n$, $B_n$, $C_n$ and $D_n$ are variable parameters. This
formula is an extension of that given in Ref.~\cite{stolo} and
includes a large-scale term. Such extensions have been attempted
earlier (e.g., Ref.~\cite{detlef}). Antonia et al.~\cite{Ant00}
have successfully tested the function (\ref{interp}). Dhruva
\cite{dhruva} has shown that the present interpolation formula
works extremely well for longitudinal structure functions of order
2, 4 and 6. To reinforce this point, we test its performance by
comparing it to the measured transverse structure function,
$\alpha = \beta =1$, $r$ in the direction 3. For each data set,
the height of the probe is assumed to be the large scale $L$. The
fit is shown for the transverse structure function of orders 2 and
4 at the 0.54 m probe in Fig.~\ref{interp_test}. The agreement
between the formula and the data is excellent. Taken together with
similar conclusions in \cite{dhruva} for longitudinal structure
functions, we conclude that the interpolation formula describes
the familiar structure functions very well. For this pragmatic
reason, we shall adopt it for our purposes here, and test the
robustness of the results obtained in the appendix C.

\begin{figure}
\centerline{\includegraphics[scale=0.4]{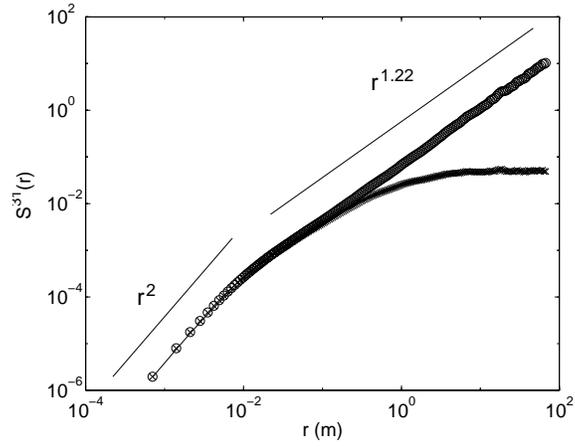}}
\caption{Log-log plot of second order mixed structure function at
0.54m. X denote data, the solid line is the interpolation fit (not
visible beyond an r of $10^{-1}$m because of the closely packed
symbols), and O correspond to the large-scale compensated
function.} \label{comp2_54}
\end{figure}
\begin{figure}
\centerline{\includegraphics[scale=0.4]{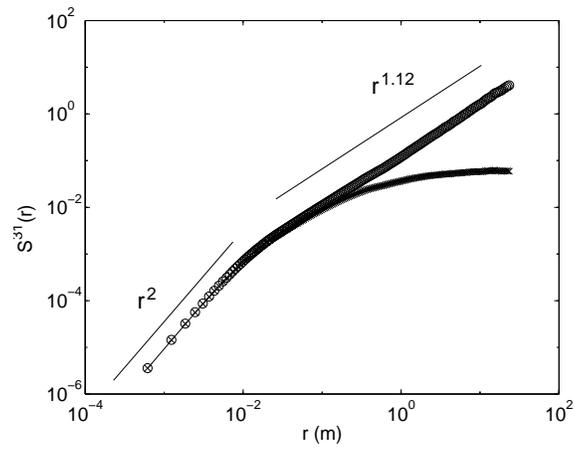}}
\caption{Log-log plot of second order mixed structure function at
0.27m. The legend is the same as for Fig. \ref{comp2_54}.}
\label{comp2_27}
\end{figure}
In the formula (\ref{interp}), the large scale behavior is given by
the factor $(1 + D_2(r/L))^{2C_2-2}$. If the measured structure
function is divided by this factor, we should recover the contribution
of the remaining parts---in particular the inertial range part, with
the leading order scaling exponent given by $2-2C_2$. Figures
\ref{comp2_54} and \ref{comp2_27} display a second-order anisotropic
structure function for two heights above the ground. Presumably
because of the finiteness of the Reynolds number and the relatively
large shear effect, the scaling in the intermediate range $\eta << r
<< L$ is not apparent. However, by dividing out the large scale
contribution as described above, we see two distinct regions of
scaling; the dissipative range of $\sim r^2$ and the extended
mid-range which scales with exponent between $1.22$ and $1.12$. The
advantage of the scaling function is thus evident: it has allowed us
to extract a scaling exponent that is most likely to be due to
anisotropy. The values of the fitted parameters and the corresponding
$\zeta_2^{(2)}$ are given in Tables \ref{results.54} and
\ref{results.27} for the probes at 0.54 m and at 0.27 m
respectively. The error on the measurement of the exponent $C_2$ at 
0.54 m is about
0.05 while at 0.27 m it is about 0.08. This gives an error on the
estimates of $\zeta_2^{(2)}$ of 0.07 and 0.11 respectively, and 
places the theoretically expected value of $\approx 4/3$ within 1.5 to
2 standard deviations of the present value. This result is consistent
with general expectations \cite{lumley} and the findings of the
previous section.

\subsubsection{Higher-order structure functions}

In general, the tensor forms contributing to the $j=0$ sector for
tensors of {\em any} rank $n$ are composed of linear combinations
of the Kronecker-$\delta$ and the components of ${\bf r}$ along the
tensor indices. The following is a list of isotropic tensor
contributions for rank 3 through 6:
\begin{itemize}

\item $n=3$: $\delta^{\alpha\beta}r^\gamma$ + permutations, and
$r^\alpha r^\beta r^\gamma$;

\item $n=4$: $\delta^{\alpha\beta}\delta^{\gamma\delta}$ + permutations,
$\delta^{\alpha\beta}r^\gamma r^\delta$ + permutations, and\\
$r^\alpha r^\beta r^\gamma r^\delta$;

\item $n=5$: $\delta^{\alpha\beta}\delta^{\gamma\delta}r^\mu$ +
permutations, $\delta^{\alpha\beta}r^\gamma r^\delta r^\mu$ +
permutations, and $r^\alpha r^\beta r^\gamma r^\delta r^\mu$;

\item $n=6$: $\delta^{\alpha\beta}\delta^{\gamma\delta}\delta^{\mu\nu}$ +
permutations, $\delta^{\alpha\beta}\delta^{\gamma\delta}r^\mu
r^\nu$ + permutations, \\
$\delta^{\alpha\beta}r^\gamma r^\delta
r^\mu r^\nu$ + permutations, and $r^\alpha r^\beta r^\gamma
r^\delta r^\mu r^\nu$.
\end{itemize}

\begin{figure}
\centerline{\includegraphics[scale=0.4]{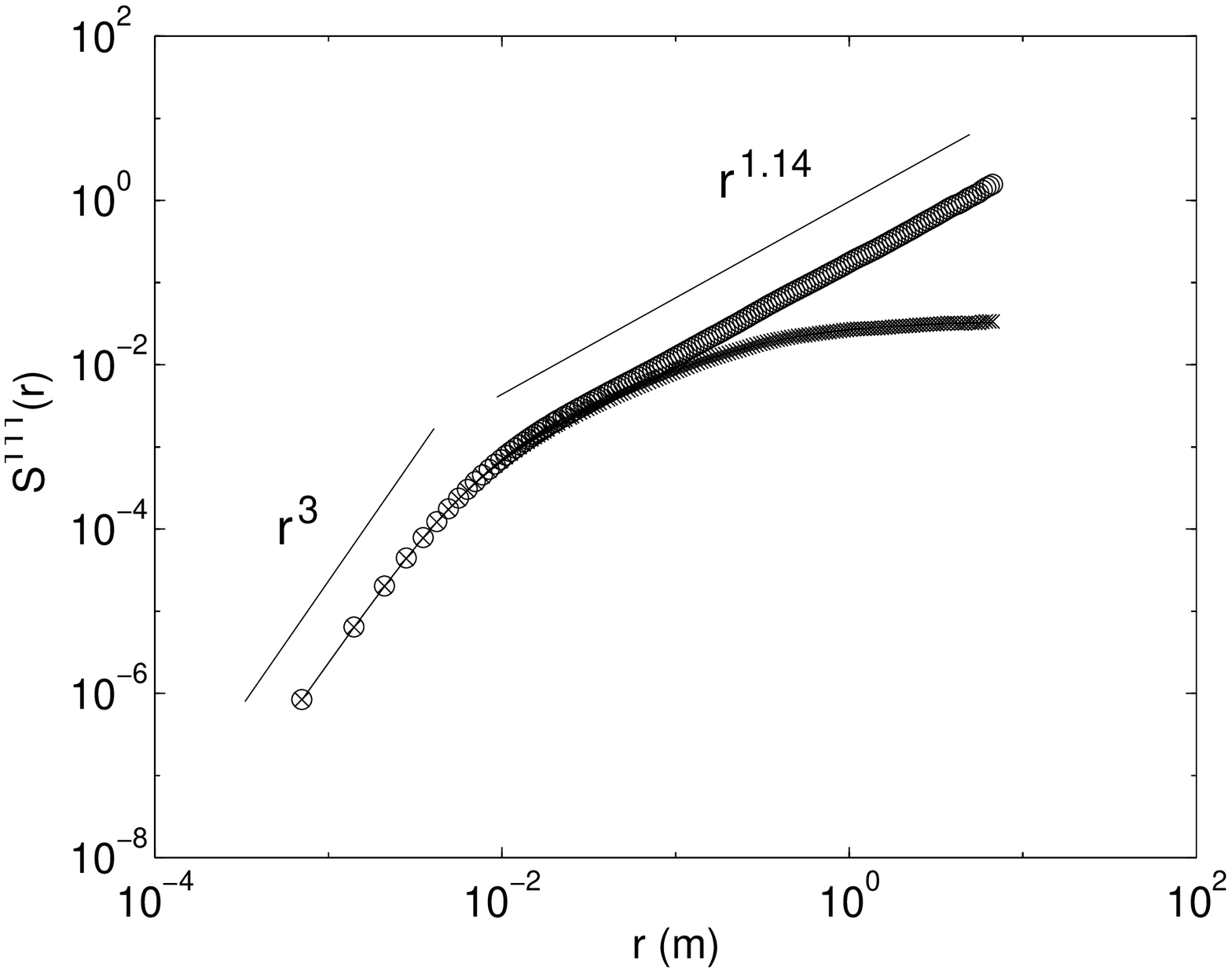}}
\caption{Log-log plot of third-order transverse structure function
at 0.54m. The legend is the same as for Fig. \ref{comp2_54}.}
\label{comp3_54}
\end{figure}

\begin{figure}
\centerline{\includegraphics[scale=0.4]{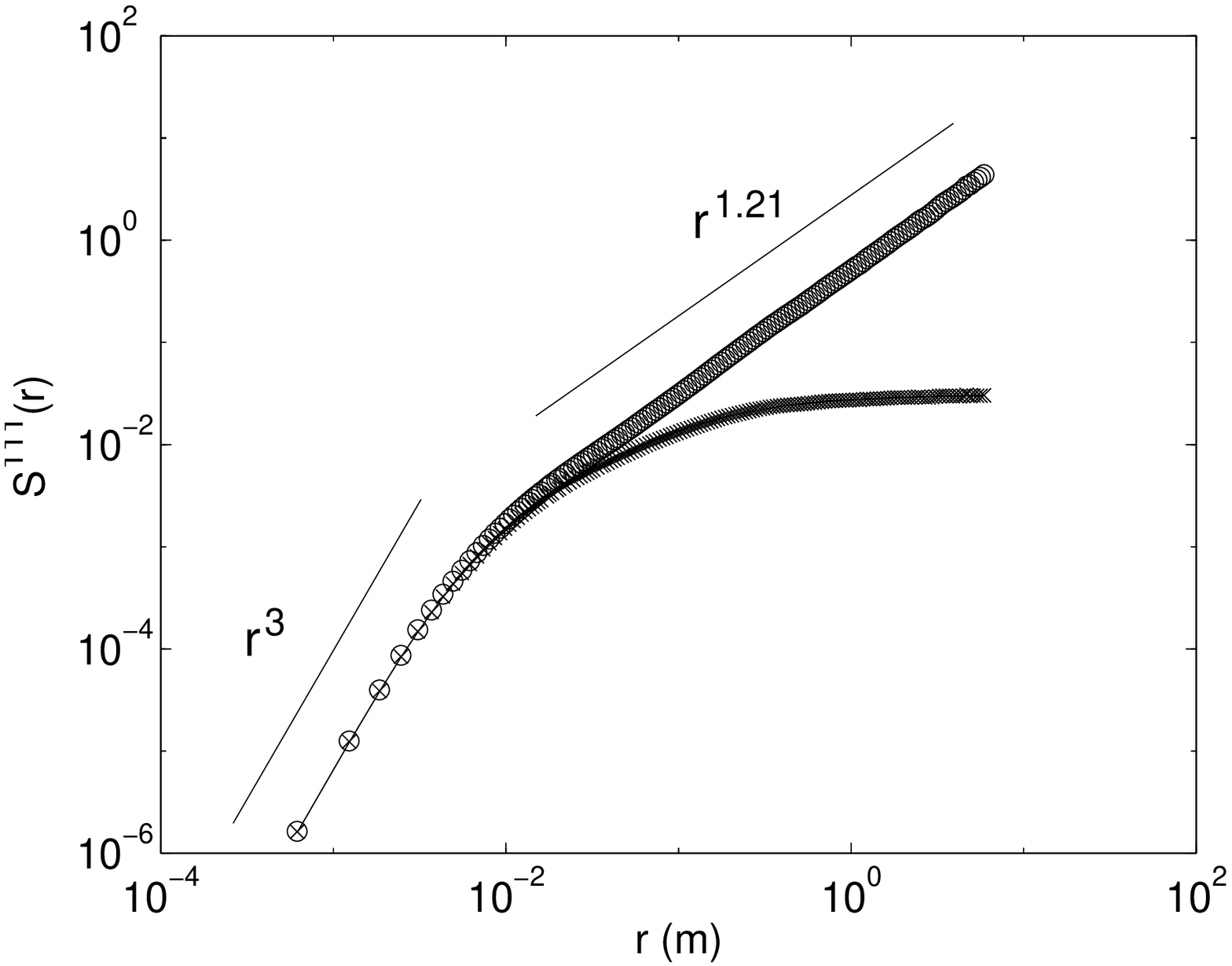}}
\caption{Log-log plot of third-order transverse structure function
at 0.27m. The legend is the same as for Fig. \ref{comp2_54}.}
\label{comp3_27}
\end{figure}

\begin{figure}
\centerline{\includegraphics[scale=0.4]{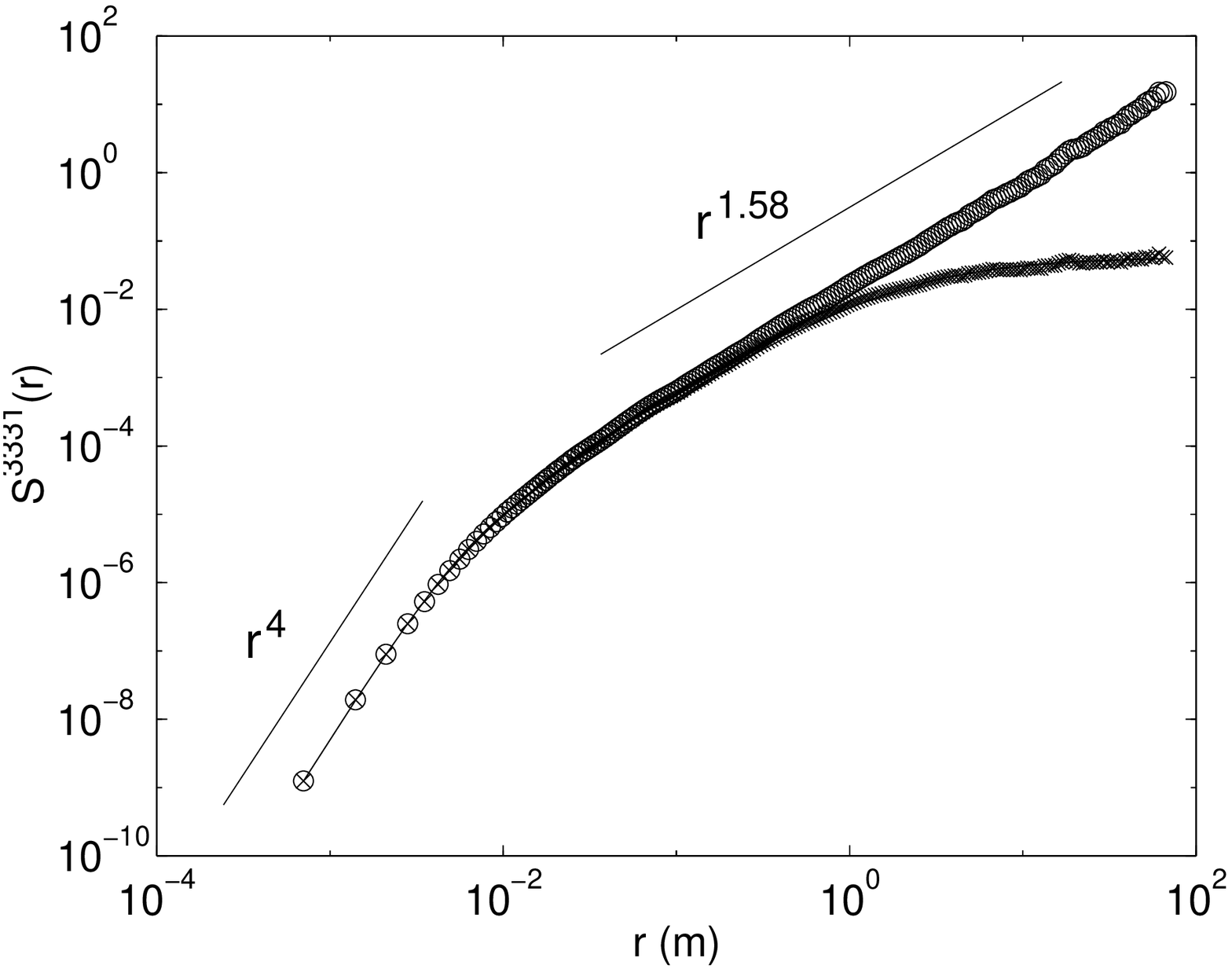}}
\caption{Log-log plot of fourth-order mixed structure function at
0.54m. The legend is the same as for Fig. \ref{comp2_54}.}
\label{comp4_54}
\end{figure}

\begin{figure}
\centerline{\includegraphics[scale=0.4]{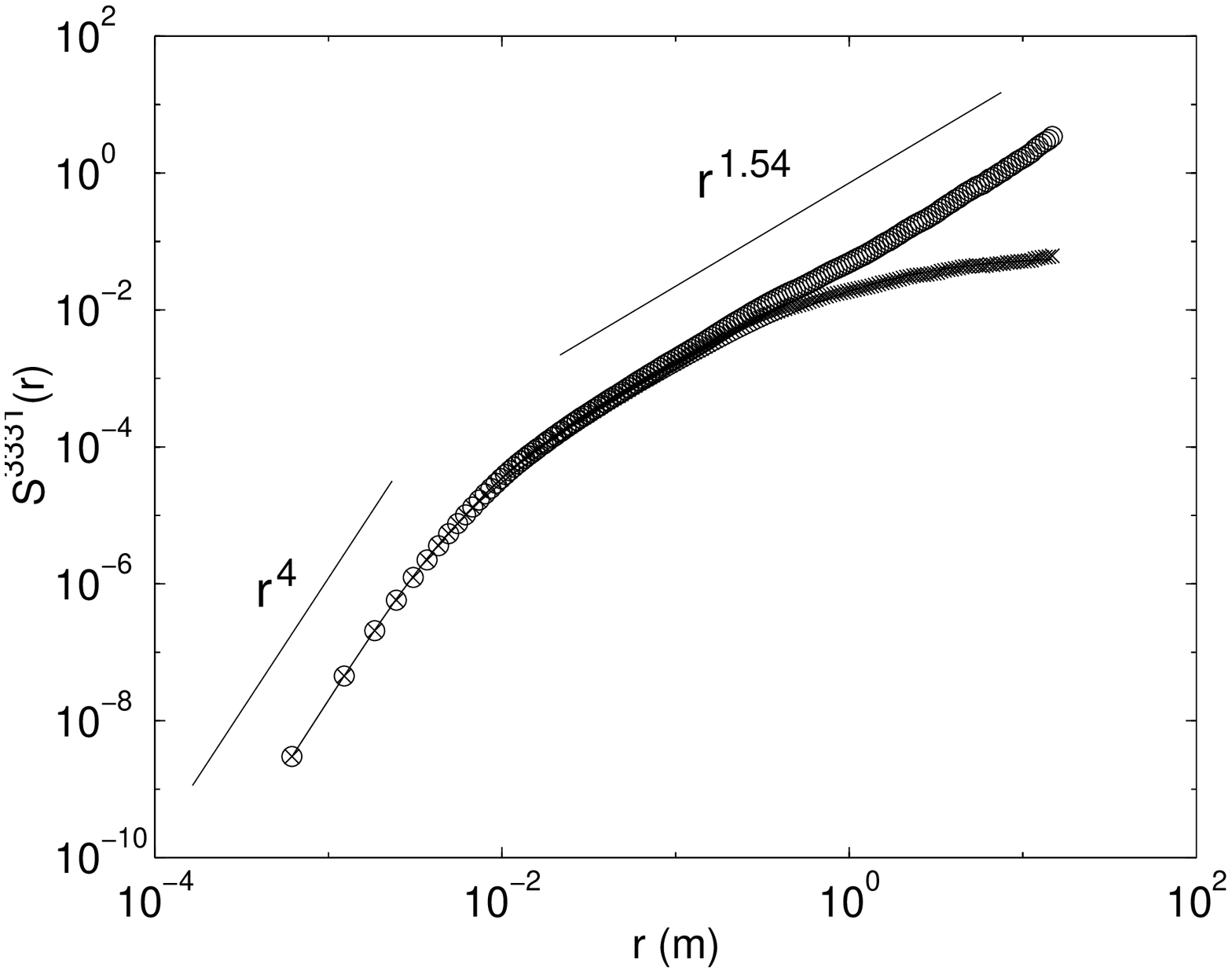}}
\caption{Log-log plot of fourth-order mixed structure function at
0.27m. The legend is the same as for Fig. \ref{comp2_54}.}
\label{comp4_27}
\end{figure}

\begin{figure}
\centerline{\includegraphics[scale=0.4]{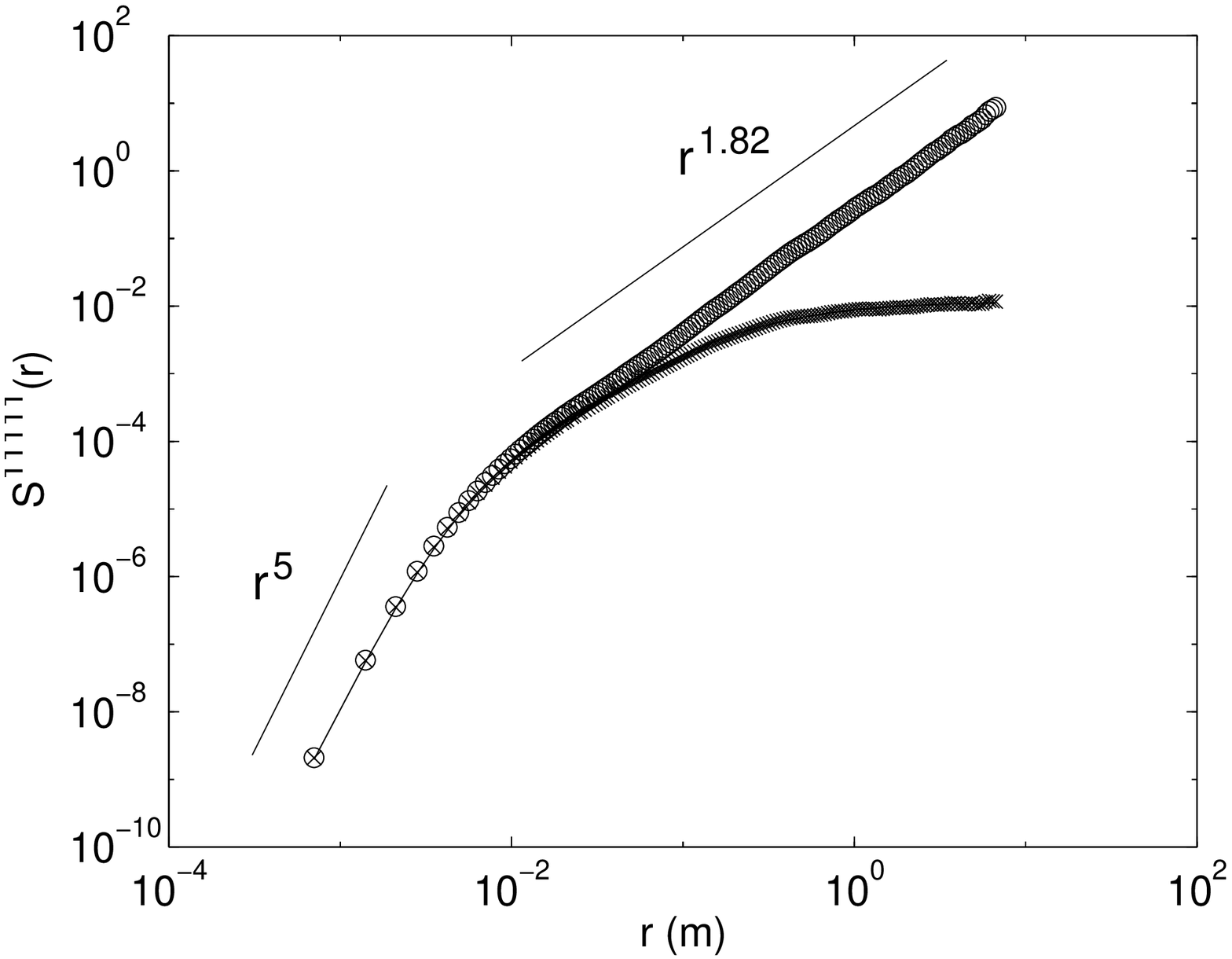}}
\caption{Log-log plot of fifth-order transverse structure function
at 0.54m. The legend is the same as for Fig. \ref{comp2_54}.}
\label{comp5_54}
\end{figure}

\begin{figure}
\centerline{\includegraphics[scale=0.4]{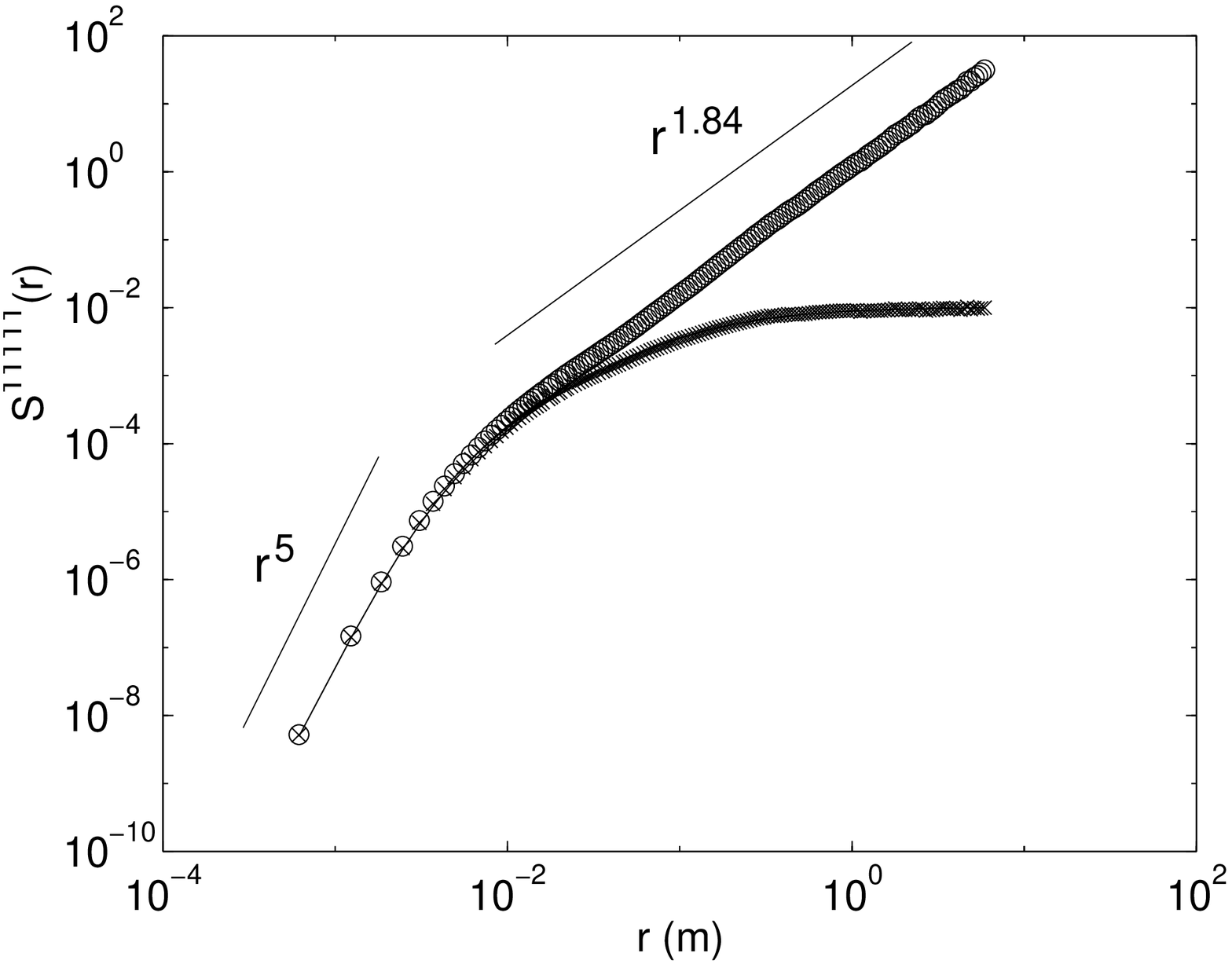}}
\caption{Log-log plot of fifth-order transverse structure function
at 0.27m. The legend is the same as for Fig. \ref{comp2_54}. }
\label{comp5_27}
\end{figure}

\begin{figure}
\centerline{\includegraphics[scale=0.4]{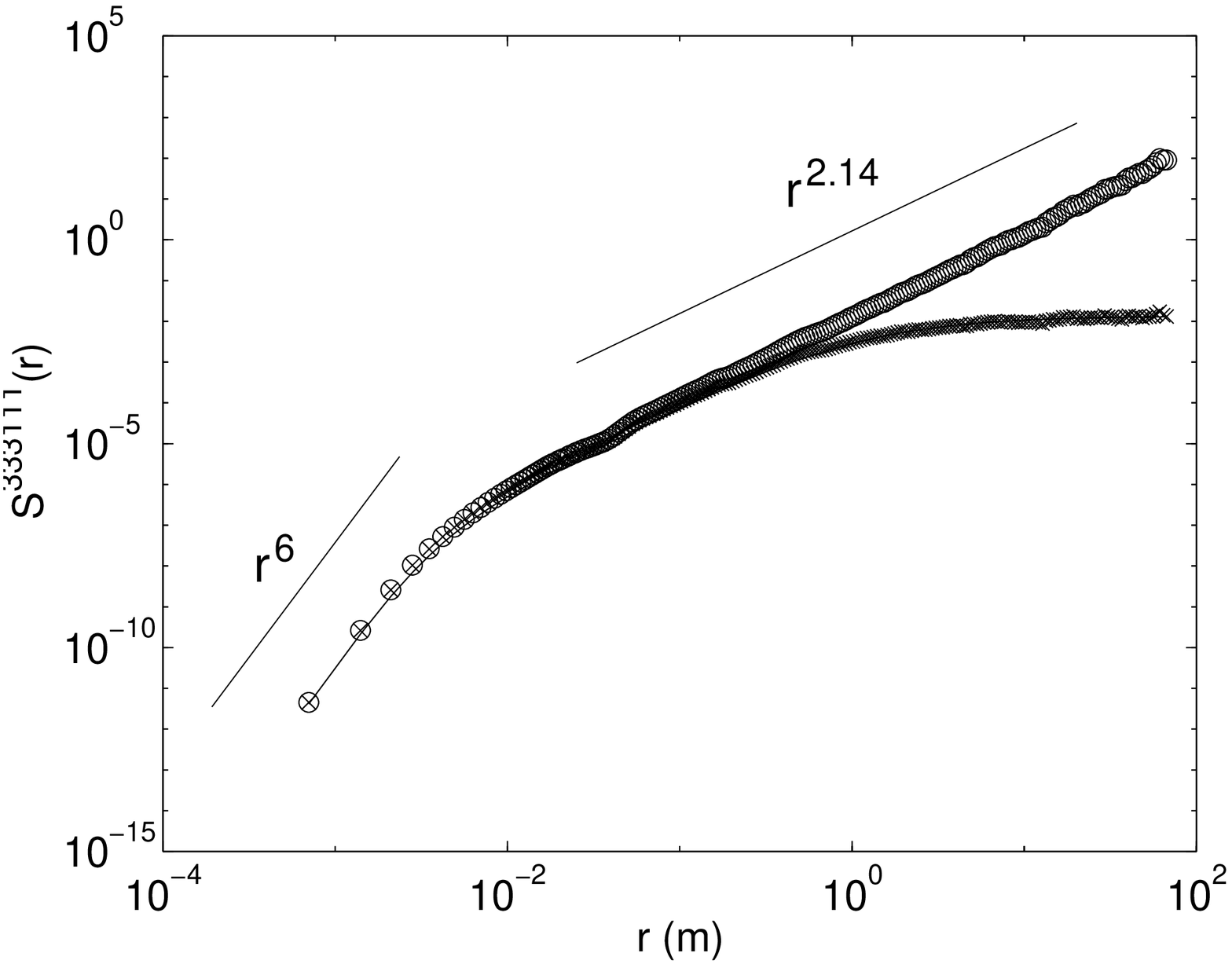}}
\caption{Log-log plot of sixth-order mixed structure function at
0.54m. The legend is the same as for Fig. \ref{comp2_54}.}
\label{comp6_54}
\end{figure}

\begin{figure}
\centerline{\includegraphics[scale=0.4]{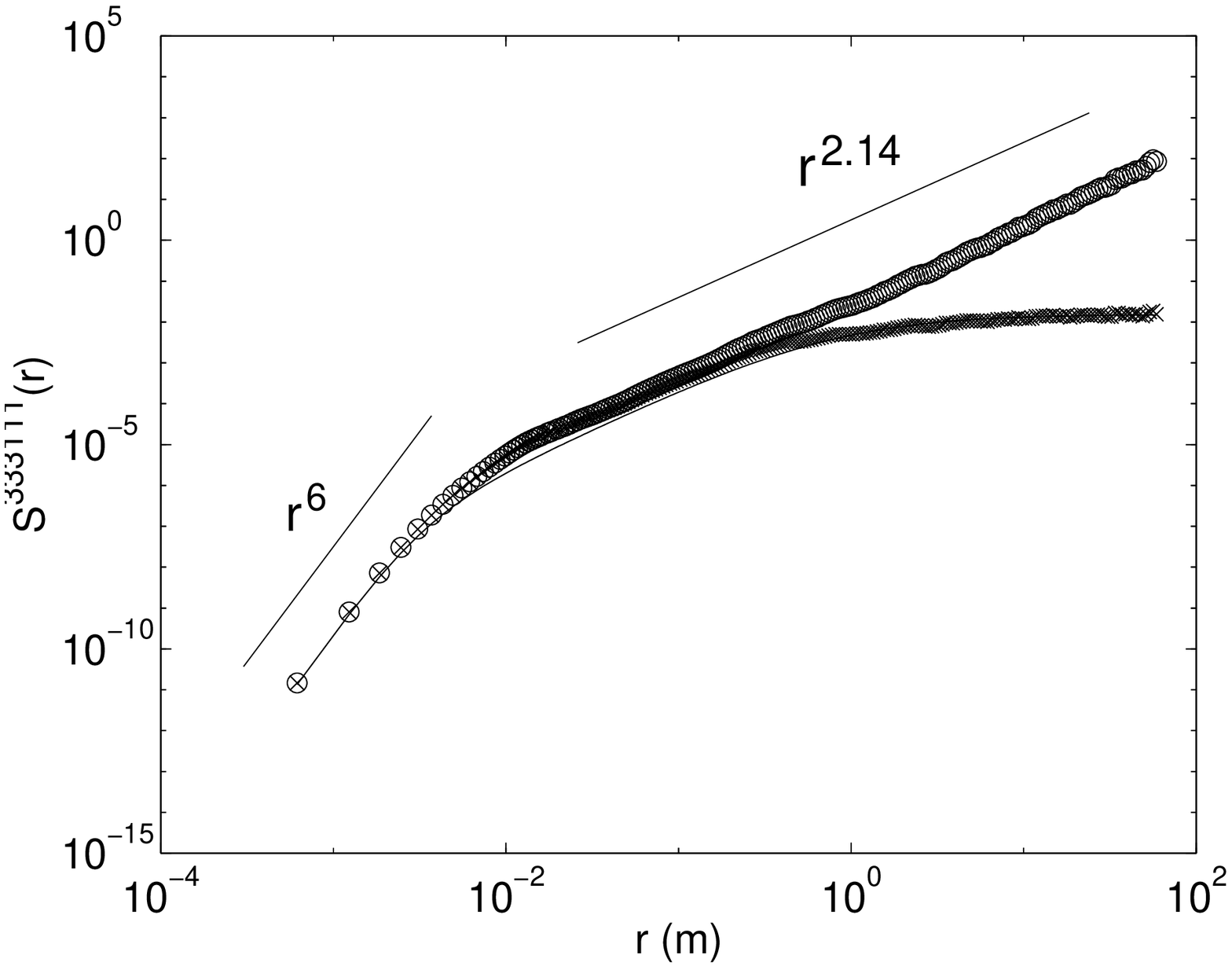}}
\caption{Log-log plot of sixth-order mixed structure function at
0.27m. The legend is the same as for Fig. \ref{comp2_54}.}
\label{comp6_27}
\end{figure}

Based on the above considerations, it can be expected that the
structure function components that are zero in the $j=0$ sector
are:
\begin{itemize}
\item $n=3$: $S^{111}$ (transverse), $S^{331}$;
\item $n=4$: $S^{3331}, S^{3111}$;
\item $n=5$: $S^{11111}$ (transverse), $S^{33111},
S^{33331}$;
\item $n=6$: $S^{333111}, S^{311111}, S^{333331}$.
\end{itemize}
Note that the odd-order transverse structure function is {\em
always} zero in the isotropic sector. The functions we shall now
consider are given in the second column of Tables \ref{results.54}
and \ref{results.27}.
\begin{table}
\centerline{\begin{tabular}{|c|c|c|c|c|c|c|c|}\hline
Order $n$ & Tensor & $A_n$ & $B_n$ & $C_n$ & $D_n$ &
$\zeta_n^{(2)}
= n - 2C_n$ & $\zeta_n$ \\ \hline
2 &$S^{31}$ & 3.9 &0.014 & 0.39 & 0.67 & 1.22 & 0.7\\
3 &$S^{111}$ &2400 & 0.010 & 0.93 & 2.28 & 1.14 & 1\\
4 &$S^{3331}$& 5200 & 0.014 & 1.21 & 0.27 & 1.58 & 1.26\\
5&$S^{11111}$ & $1.22\times10^7$& 0.029 & 1.59 & 3.09 & 1.82 &1.56 \\
6 &$S^{333111}$ & $3.75\times10^7$& 0.041 & 1.93 &0.50 & 2.14 & 1.71\\
\hline
\end{tabular}}
\caption{Structure function calculated and the anisotropic scaling
exponents for the data at 0.54m.}\label{results.54}
\end{table}
\begin{table}
\centerline{\begin{tabular}{|c|c|c|c|c|c|c|c|}\hline
Order $n$ & Tensor & $A_n$ & $B_n$ & $C_n$ & $D_n$ &
$\zeta_n^{(2)}= n - 2C_n$ & $\zeta_n$ \\ \hline
2 &$S^{31}$ & 9.4&0.005 & 0.44 & 0.52 & 1.12 & 0.7\\
3 &$S^{111}$ &6940 &0.015 & 0.89 & 2.78 & 1.21 & 1\\
4 &$S^{3331}$& $2.1\times10^4$ &0.014 & 1.23 & 0.23 & 1.54 & 1.26\\
5&$S^{11111}$ & $5.9\times10^7$ & 0.028 & 1.58 & 3.52 & 1.84 &1.56\\
6 &$S^{333111}$ & $2.7\times10^8$ & 0.038 & 2.00 &0.34 & 2.00 & 1.71\\
\hline
\end{tabular}}
\caption{Structure function calculated and the anisotropic scaling
exponents for the data at 0.27m.}\label{results.27}
\end{table}
For the case of the third and fifth order transverse structure
functions we use the moments of the {\em absolute value} of the
velocity differences in order to obtain better convergence. In
using the interpolation function we assume that the inertial range
scaling of these anisotropic components is given by a single
exponent $\zeta_n^{(j)}$ where the superscript denotes an
isotropic exponent without reference to the precise $j$ sector.
The compensated functions (with large-scale effects removed) are
shown in Figs.~\ref{comp3_54}--\ref{comp6_27}. The errors on the
value of $\zeta_n^{(2)}$ obtained are about $7\%$ at 0.54 m and
about $9\%$ at 0.27 m. For comparison, the last column in Tables
\ref{results.54} and \ref{results.27} gives the empirical values
of the isotropic scaling exponent of the same order as given in
\cite{dhruva}. The entries in this column are measurably smaller
than the corresponding non-isotropic exponents. This suggests that
the isotropic component alone survives at very small scales.

\subsection{Summary}

We have presented a new method of extracting anisotropic
exponents that avoids mixing with contributions from the isotropic
sector. We do this by explicitly constructing those tensors that
are zero in the isotropic sector. An operational step in the
extraction of the scaling exponents is the use of an interpolation
formula in the spirit of a ``scaling function''. This method has
allowed us to examine anisotropic effects in structure function
tensors of order greater than 2. The resulting
anisotropic exponents are consistently larger than those known for
isotropic parts at all orders. This strongly suggests that
anisotropy effects decrease with decreasing scale. However, the
rate of decrease is much slower than expected from dimensional
arguments (which yield 4/3, 5/3, 2, 7/3 and 8/3 for orders 2
through 6, to be compared with the values obtained at $0.54$~m of
1.22, 1.14, 1.58, 1.82, 2.14).

Our conclusions are based on the use of the interpolation formula,
given by Eq.~(\ref{interp}). While this formula is not based on a solid
theoretical framework, we have shown that it works very well in
describing the measured structure functions. We have also
performed tests of its robustness by fitting it to smaller
sections of the data in order to detect changes in the exponent. A
discussion of these checks and their results is presented in
appendix C. To the lowest order, the results are independent of
the $r$-segment to which the formula is fitted (except, of course,
when the fit is entirely for the dissipation range or in the
large-scale range). Any other formula that works equally well will
yield similar results. Even so, the formula is empirical, which is
why we have not paid much attention to the fact that the scaling
exponents obtained for the two probe positions are slightly
different, and that the second-order exponent for 0.54 m is
slightly larger than that obtained for the third-order. On the
whole, the trend is that the anisotropic exponents become larger
for larger orders of the structure function.

An interesting conclusion of the present work is that the effects
of anisotropy vanish with decreasing scale more slowly than
expected. The expectation in the light of the SO(3) formalism is
that a hierarchy of increasingly larger exponents, corresponding
to increasingly higher-order anisotropic sectors \cite{97LPP},
would exist. This expectation appears to be true in the
case of the passively advected vector field \cite{ABP00} where a
discrete spectrum of anisotropic scaling exponents is obtained
theoretically for all anisotropic sectors. In the present
experiments, the fact that anisotropic effects can be fitted
reasonably well by power laws suggests that the high-order effects
may be small. It is perhaps true, however, that the power-laws
described here may contain high-order corrections, and that the
exponents deduced for the behavior of anisotropy may indeed
undergo some revision when contributions from other sectors of the
SO(3) decomposition are also considered. In spite of this
possibility, we wish to emphasize that the anisotropy effects {\it for
each order} of the structure function appear to be well described
by something close to a power law with a single exponent. The
magnitudes of the anisotropic exponents in each order indicate
that the roll-off from isotropy happens less sharply than
previously thought, but the roll-off occurs nevertheless. The
higher-order objects considered here have not been studied
extensively in the light of anisotropy.
\section{Conclusions}
We have used the SO(3) decomposition of tensorial objects such as
structure functions measured in fluid turbulence experiments. This
enables us to separate, for any given order structure function,
the isotropic part from the anisotropic part. We have used
experimental data at very high Reynolds numbers (Taylor microscale
Reynolds numbers of up to 20,000) to carry through this
decomposition.

We considered the lowest order anisotropic contribution to the
second order tensor functions of velocity in the atmospheric
boundary layer. For second order structure functions, we have
extracted the anisotropic scaling exponent from the experimental
data. We also used the full form of the $j=2$ tensor, as well as
the axisymmetric version of it. In both instances, we have shown
that the $j=2$ exponent $\zeta^{(2)}_2 \approx 1.36$. This
compares favorably with the estimate based on purely dimensional
grounds. It should be noted that the dimensional estimates do not
take into account possible anomalous scaling. We expect that
turbulence flows will exhibit anomalous scaling in the isotropic
sector, and in every anisotropic sector in the SO(3) hierarchy. 
By suitably arranging the measurement configuration, we have also
extracted the $j=1$ contribution, and find the appropriate scaling
exponent to be close to the dimensional estimate of unity.

For high orders, we have presented a new method for the extraction
of anisotropic exponents that avoids mixing with contributions
from the isotropic sector. We do this by explicitly constructing
those tensors that are zero in the isotropic sector. The resulting
anisotropic exponents are consistently larger than those known for
isotropic parts at all orders. This strongly suggests that
anisotropy effects decrease with decreasing scale. However, the
rate of decrease is slower than expected from dimensional
arguments (which yield 4/3, 5/3, 2, 7/3 and 8/3 for orders 2
through 6, to be compared with the values obtained at $0.54$~m of
1.22, 1.14, 1.58, 1.82, 2.14).

Armed with these details, it is now possible to say something of
value about local isotropy. These statements are interspersed
through out the text, but the most important conclusion is that
the effects of anisotropy within a structure function of a given
order do vanish with decreasing scale, though more slowly than
expected. This provides a rich perspective on the notion of local
isotropy and removes a hurdle towards the search for a universal
theory of small-scale turbulence.
\appendix        
\section{Full form for the $j=2$ contribution for the homogeneous case}
Each index $j$ in the SO(3) decomposition of an $n$-rank tensor
labels a $2j+1$ dimensional SO(3) representation. Each dimension
is labeled by $m=-j,-j+1,\dots j$. The $j=0$ sector is the
isotropic contribution while higher order $j$'s should describe
any anisotropy. The $j=0$ terms are well known to be
\begin{equation} S_{j=0}^{\alpha\beta}({\bf
r})=c_0r^{\zeta_2} \left[(2+\zeta_2) \delta^
{\alpha\beta}-\zeta_2{r^\alpha r^\beta\over r^2}\right]\,
\label{Siso}
\end{equation} where $\zeta_2\approx0.69$ is the known universal scaling
exponent for the isotropic contribution, and $c_0$ is an unknown
coefficient that depends on the boundary conditions of the flow.
For the $j=2$ sector, which is the lowest contribution to
anisotropy to the homogeneous structure function, the $m=0$
(axisymmetric) terms were derived from constraints of symmetry,
even parity (because of homogeneity) and incompressibility on the
second order structure function to be \cite{98ADKLPS}
\begin{eqnarray}\label{m0}
&&S^{\alpha\beta}_{j=2,m=0}({\bf r}) =
ar^{\zeta_2^{(2)}}\Big[(\zeta_2^{(2)}
-2)\delta^{\alpha\beta} -
\zeta_2^{(2)}(\zeta_2^{(2)}+6)\nonumber\\ &\times&\delta^{\alpha\beta}
{({\bf n}\cdot {\bf r})^2 \over r^2}+2\zeta_2^{(2)}(\zeta_2^{(2)}-2){r^\alpha
r^\beta({\bf n}\cdot {\bf r})^2 \over r^4}\nonumber\\
&+&([\zeta_2^{(2)}]^2+3\zeta_2^{(2)}+6)n^\alpha n^\beta \nonumber\\
&-&{\zeta_2^{(2)}(\zeta_2^{(2)}-2)\over r^2}(r^\alpha n^\beta + r^\beta
n^\alpha)({\bf n}\cdot {\bf r})\Big]\label{final}\\ &+&
br^{\zeta_2^{(2)}}\Big [-(\zeta_2^{(2)} +3)(\zeta_2^{(2)}+2)\delta^
{\alpha\beta}({\bf n}\cdot {\bf r})^2 +
{r^\alpha r^\beta \over r^2} \nonumber\\ &+& (\zeta_2^{(2)} +3)
(\zeta_2^{(2)}+2)n^\alpha n^\beta + (2\zeta_2^ {(2)}+1)(\zeta_2^{(2)}-2)
\nonumber\\
&\times&{{r^\alpha}{r^\beta}{({\bf n}\cdot {\bf r})^2} \over r^4}-
([\zeta_2^{(2)}]^2 - 4)(r^\alpha n^\beta + r^\beta n^\alpha)({\bf
n}\cdot {\bf r})\Big]. \nonumber
\end{eqnarray}
Here $\zeta_2^{(2)}$ is the universal scaling exponent for the
$j=2$ anisotropic sector and $a$ and $b$ are independent unknown
coefficients to be determined by the boundary conditions. We would
now like to derive the remaining $m=\pm1$, and $m=\pm2$ components
\begin{equation}
S_{j=2,m}^{\alpha\beta}=
    \sum_q {a_{q,2,m}r^{\zeta_2^{(2)}}B_{q,2,m}^{\alpha\beta} (\bf
{\hat r})},
\end{equation}
where $\zeta_2^{2}$ is the scaling exponent of the $j=2$ SO(3)
representation of the $n=2$ rank correlation function. The
$B_{q,j,m}^{\alpha\beta}(\bf  {\hat r})$ are the basis functions
in the SO(3) representation of the structure function. The $q$
label denotes the different possible ways of arriving at the the
same $j$ and runs over all such terms with the same parity and
symmetry (a consequence of homogeneity and hence the constraint of
incompressibility); see Ref.~\cite{99ALP}. In our case, even
parity and symmetric in the two indices. In all that follows, we
work closely with the procedure outlined in \cite{99ALP}.
Following the convention in \cite{99ALP} the $q$'s to sum over are
$q=\{1,7,9,5\}$. The incompressibility condition $\partial_\alpha
u^\alpha = 0$ coupled with homogeneity can be used to give
relations between the $a_{q,j,m}$ for a given $(j,m)$. That is,
for $j=2$, $m=-2\dots 2$, we have
\begin{eqnarray} (\zeta_2^{(2)} - 2)a_{1,2,m} +
2(\zeta_2^{(2)} - 2)a_{7,2,m} + (\zeta_2^{(2)} + 2)a_{9,2,m} &=&
0 \nonumber\\
 a_{1,2,m} + (\zeta_2^{(2)} + 3)a_{7,2,m} +
\zeta_2^{(2)}a_{5,2,m} &=& 0.
\end{eqnarray}

We solve equations (3) in order to obtain $a_{5,2,m}$ and
$a_{7,2,m}$ in terms of linear combinations of $a_{1,2,m}$ and
$a_{9,2,m}$:
\begin{eqnarray}
a_{5,2,m} &=& {a_{1,2,m}([\zeta_2^{(2)}]^2 - \zeta_2^{(2)} - 2) + a_{9,2,m}
([\zeta_2^{(2)}]^2 + 5 \zeta_2^{(2)} + 6) \over
2\zeta_2^{(2)}(\zeta_2^{(2)} - 2)}\nonumber\\
a_{7,2,m} &=& {a_{1,2,m}(2-\zeta_2^{(2)}) -
a_{9,2,m}(2+\zeta_2^{(2)}) \over 2(\zeta_2^{(2)} - 2)}.
\end{eqnarray}

Using the above constraints on the coefficients, we are now left
with a linear combination of just two linearly independent tensor
forms {\em for each m}:
\begin{eqnarray}\label{genl-s2m}
S^{\alpha\beta}_{j=2,m} &=&
a_{9,2,m}r^{\zeta_2^{(2)}}[-\zeta_2^{(2)}(2+\zeta_2^{(2)})
B_{7,2,m}^{\alpha\beta}({\bf  {\hat r}}) \nonumber\\
&&\mbox{~}\hspace{1.5cm}+ 2\zeta_2^{(2)}(\zeta_2^{(2)} - 2)
B_{9,2,m}^{\alpha\beta}({\bf  {\hat r}}) \nonumber \\
&&\mbox{~}\hspace{1.5cm}+([\zeta_2^{(2)}]^2+5\zeta_2^{(2)}+6)
B_{5,2,m}^{\alpha\beta}({\bf {\hat r}})] \nonumber\\
&+&a_{1,2,m}r^{\zeta_2^{(2)}}[2\zeta_2^{(2)}(\zeta_2^{(2)} - 2)
B_{1,2,m}^{\alpha\beta}({\bf  {\hat r}}) \nonumber\\
&&\mbox{~}\hspace{1.5cm}-\zeta_2^{(2)}(\zeta_2^{(2)}-2)
B_{7,2,m}^{\alpha\beta}({\bf  {\hat r}})\\
&&\mbox{~}\hspace{1.5cm}+([\zeta_2^{(2)}]^2-\zeta_2^{(2)}-2)
B_{5,2,m}^{\alpha\beta}({\bf {\hat r}})].\nonumber
\end{eqnarray}

The task remains to find the explicit form of the basis tensor functions
$B_{q,2,m}^{\alpha\beta}({\bf  {\hat r}})$, $q\in\{1,7,9,5\}$,
$m\in\{\pm1,\pm2\}$
\\
$\bullet$\,$ B_{1,2,m}^{\alpha\beta}({\bf  {\hat r}}) \equiv
r^{-2}\delta^{\alpha\beta}r^j Y_{2m}({\bf  {\hat r}})$ \\
$\bullet$\,$ B_{7,2,m}^{\alpha\beta}({\bf  {\hat r}}) \equiv r^{-2}[r^\alpha
\partial^\beta +
r^\beta \partial^\alpha]r^2 Y_{2m}({\bf  {\hat r}})$ \\
$\bullet$\,$ B_{9,2,m}^{\alpha\beta}({\bf  {\hat r}}) \equiv r^{-4}r^\alpha
r^\beta r^2 Y_{2m}({\bf  {\hat r}})$ \\
$\bullet$\,$ B_{5,2,m}^{\alpha\beta}({\bf  {\hat r}}) \equiv \partial^\alpha
\partial^\beta r^2 Y_{jm}({\bf  {\hat r}}).$

We obtain the $m=\{\pm1,\pm2\}$ basis functions in the following derivation.
We first note that it is more convenient to form a real basis from the $
r^2 Y_{2m}({\bf  {\hat r}})$ since we ultimately wish
to fit to real quantities and extract real best-fit parameters.
We therefore form the $r^2 {\widetilde Y}_{2k}({\bf  {\hat r}})$ ($k=
-1,0,1$) as follows:
\begin{eqnarray}
r^2 {\widetilde Y}_{2\; 0}(\hat {\bf  r})&=&r^2 Y_{2\;0}(\hat {\bf  r})
=r^2 \cos^2 \theta = r{_3}^2\nonumber \\
r^2 {\widetilde Y}_{2\;-1}(\hat{\bf  r}) &=&r^2 {Y_{2\;-1}(\hat {\bf  r})
-Y_{2\;+1}(\hat {\bf  r}) \over 2}\nonumber\\
&=&{r^2 \over 2}\Big[(\cos\phi -i\sin\phi)\cos\theta\sin\theta \nonumber\\
&&\hspace{0.7cm}+(\cos\phi + i\sin\phi)\cos\theta\sin\theta \Big]
\nonumber\\
&=&r^2\cos\theta\sin\theta\cos\phi = r_3 r_1 \nonumber\\
r^2 {\widetilde Y}_{2\;+1}(\hat {\bf  r}) &=&r^2 {Y_{2\; -1}(\hat {\bf r})
+ Y_{2\; +1}(\hat{\bf  r}) \over -2i} \nonumber\\
&=&{r^2 \over -2i}\Big[(\cos\phi-i\sin\phi)\cos\theta\sin\theta\nonumber\\
&&\hspace{0.7cm}-(\cos\phi + i\sin\phi)\cos\theta\sin\theta \Big]
\nonumber\\
&=&r^2\cos\theta\sin\theta\sin\phi = r_3 r_2\nonumber\\
r^2 {\widetilde Y}_{2\;-2}(\hat {\bf  r}) &=& r^2 {Y_{2\; 2}(\hat {\bf r})
-Y_{2\; -2}(\hat {\bf  r}) \over 2i} \nonumber\\
&=& {r^2\over 2i}\Big[(\cos2\phi +i\sin2\phi)\sin^2\theta - (\cos2\phi -
i\sin2\phi)\sin^2\theta \Big]\nonumber \\
&=& r^2\sin2\phi\sin^2\theta = 2r_1 r_2 \nonumber \\
r^2 {\widetilde Y}_{2\;+2}(\hat {\bf  r}) &=& r^2 {Y_{2\; 2}(\hat{\bf  r})
+ Y_{2\; -2}(\hat{\bf  r}) \over 2} \nonumber\\
&=& {r^2\over2}\Big[(\cos2\phi +i\sin2\phi)\sin^2\theta + (\cos2\phi - i\sin2\phi)\sin^2\theta\Big]\nonumber \\
&=& r^2 \cos2\phi\sin^2\theta = r_1^2 - r_2^2.
\end{eqnarray}
This new basis of $r^2 {\tilde Y}_{2k}{(\bf  r)}$ is equivalent to
the $r^2 Y_{jm}{(\bf  r)}$ themselves as they form a complete,
orthogonal set (in the new k's). We omit the normalization
constants for the spherical harmonics for notational convenience.
The subscripts on $r$ denote its components along the 1 ($m$), 2
($p$) and 3 ($n$) directions; ${\bf  m}$ denotes the shear
direction, ${\bf  p}$ the horizontal direction parallel to the
boundary and orthogonal to the mean wind direction and ${\bf  n}$
the direction of the mean wind. This notation simplifies the
derivatives when we form the different basis tensors; the only
thing to remember is that
\begin{eqnarray}
\partial^\alpha r_1 = \partial^\alpha (\bf  {r \cdot m}) = m^\alpha
\nonumber\\ \partial^\alpha r_2 = \partial^\alpha (\bf  {r \cdot
p})= p^\alpha \nonumber\\ \partial^\alpha r_3 = \partial^\alpha
(\bf  {r \cdot n}) = n^\alpha. \end{eqnarray}

We use the above identities to proceed to derive the basis tensor functions
\begin{eqnarray}
B_{1,2, -1}^{\alpha\beta}({\bf  {\hat r}}) &=&
r^{-2}\delta^{\alpha\beta} ({\bf  r \cdot n})({\bf  r \cdot m})
\nonumber\\ B_{7,2, -1}^{\alpha\beta} ({\bf  {\hat r}}) &=&
r^{-2}[(r^\alpha m^\beta + r^\beta m^\alpha)({\bf  r \cdot n}) +
(r^\alpha n^\beta + r^\beta n^\alpha)({\bf  r \cdot m})]
\nonumber\\ B_{9,2, -1}^{\alpha\beta}({\bf {\hat r}}) &=&r^{-2}
r^\alpha r^\beta ({\bf  r \cdot n})({\bf  r \cdot m})\nonumber\\
B_{5,2, -1}^{\alpha\beta}({\bf  {\hat r}}) &=&n^\alpha m^\beta +
n^\beta m^\alpha \nonumber\\ B_{1,2, 1}^{\alpha\beta}({\bf  {\hat
r}}) &=& r^{-2}\delta^{\alpha\beta}({\bf r \cdot n})({\bf  r \cdot
p})\nonumber\\ B_{7,2, 1}^{\alpha\beta}({\bf  {\hat r}}) &=&
r^{-2}[(r^\alpha p^\beta + r^\beta p^\alpha)({\bf  r \cdot n}) +
(r^\alpha n^\beta + r^\beta n^\alpha)({\bf  r \cdot p})]
\nonumber\\ B_{9,2, 1}^{\alpha\beta}({\bf {\hat r}}) &=&r^{-2}
r^\alpha r^\beta ({\bf  r \cdot n})({\bf  r \cdot p})\nonumber\\
B_{5,2,1}^{\alpha\beta}({\bf  {\hat r}}) &=& n^\alpha p^\beta +
n^\beta p^\alpha \nonumber\\ B_{1,2, -2}^{\alpha\beta}({\bf  {\hat
r}}) &=&2 r^{-2}\delta^{\alpha\beta}({\bf r \cdot m})({\bf  r
\cdot p})\nonumber\\ B_{7,2, -2}^{\alpha\beta}({\bf {\hat r}})
&=&2 r^{-2}[(r^\alpha p^\beta + r^\beta p^\alpha)({\bf  r \cdot
m}) + (r^\alpha m^\beta + r^\beta m^\alpha)({\bf  r \cdot
p})]\nonumber \\ B_{9,2, -2}^{\alpha\beta}({\bf {\hat r}}) &=& 2
r^{-2} r^\alpha r^\beta ({\bf  r \cdot m})({\bf  r \cdot
p})\nonumber\\ B_{5,2,-2}^{\alpha\beta}({\bf  {\hat r}}) &=& 2
(m^\alpha p^\beta + m^\beta p^\alpha) \nonumber\\ B_{1,2,
2}^{\alpha\beta}({\bf  {\hat r}}) &=&
r^{-2}\delta^{\alpha\beta}[({\bf  r \cdot m})^2 - ({\bf  r \cdot
p})^2] \nonumber\\ B_{7,2, 2}^{\alpha\beta}({\bf  {\hat r}}) &=&2
r^{-2}[(r^\alpha m^\beta + r^\beta m^\alpha)({\bf  r \cdot m}) -
(r^\alpha p^\beta + r^\beta p^\alpha)({\bf  r \cdot p})]\nonumber
\\ B_{9,2, 2}^{\alpha\beta}({\bf  {\hat r}}) &=& r^{-2} r^\alpha
r^\beta [({\bf  r \cdot m})^2 - ({\bf  r \cdot p})^2] \nonumber\\
B_{5,2,2}^{\alpha\beta}({\bf  {\hat r}}) &=& 2 (m^\alpha m^\beta -
p^\alpha p^\beta). \end{eqnarray}

Substituting these tensors forms into Eq.~\ref{genl-s2m} we obtain
the full tensor forms for the $j=2$ non-axisymmetric terms, with
two independent coefficients for each k:
\begin{eqnarray}\label{ktens}
S^{\alpha\beta}_{j=2,k=-1}({\bf  r})
&=&a_{9,2,-1}r^{\zeta_2^{(2)}}{\Big [}([\zeta_2^{(2)}]^2+5\zeta_2^{(2)}+6)(n^\alpha m^\beta + n^\beta
m^\alpha) \nonumber\\
&+& 2\zeta_2^{(2)}(\zeta_2^{(2)} - 2)
r^{-4} r^\alpha r^\beta ({\bf  r \cdot n})({\bf  r \cdot m})\nonumber\\
&-&\zeta_2^{(2)}(2+\zeta_2^{(2)})
r^{-2}[(r^\alpha m^\beta + r^\beta m^\alpha)({\bf  r \cdot n})\nonumber\\
&&~\hspace{2.5cm}+(r^\alpha n^\beta + r^\beta n^\alpha)({\bf r \cdot m})]{\Big ]}\nonumber\\
&+&a_{1,2,-1}r^{\zeta_2^{(2)}}{\Big [}([\zeta_2^{(2)}]^2-\zeta_2^{(2)}-2)(n^\alpha m^\beta + n^\beta m^\alpha)\nonumber\\
&+&2\zeta_2^{(2)}(\zeta_2^{(2)} - 2)
r^{-2}\delta^{\alpha\beta}({\bf  r \cdot n})({\bf  r \cdot m}) \nonumber\\
&-& \zeta_2^{(2)}(\zeta_2^{(2)}-2) r^{-2}[(r^\alpha m^\beta + r^\beta
m^\alpha)({\bf  r \cdot n})\nonumber\\
&&~\hspace{2.5cm}+(r^\alpha n^\beta + r^\beta n^\alpha)({\bf  r \cdot m})]{\Big ]} \nonumber\\
S^{\alpha\beta}_{j=2,k=1}({\bf  r})
&=&a_{9,2,1}r^{\zeta_2^{(2)}}{\Big [}([\zeta_2^{(2)}]^2+5\zeta_2^{(2)}+6)
(n^\alpha p^\beta + n^\beta p^\alpha)\nonumber\\
&+&2\zeta_2^{(2)}(\zeta_2^{(2)} - 2)
r^{-4} r^\alpha r^\beta ({\bf  r \cdot n})({\bf  r \cdot p})\nonumber\\
&-&\zeta_2^{(2)}(2+\zeta_2^{(2)})
r^{-2}[(r^\alpha p^\beta + r^\beta p^\alpha)({\bf  r \cdot n})\nonumber\\
&&~\hspace{2.5cm}+(r^\alpha n^\beta + r^\beta n^\alpha)({\bf r\cdot p})]
{\Big ]}\nonumber\\
&+&a_{1,2,1}r^{\zeta_2^{(2)}}{\Big [}([\zeta_2^{(2)}]^2-\zeta_2^{(2)}-2)(n^\alpha p^\beta + n^\beta p^\alpha)\nonumber\\
&+&2\zeta_2^{(2)}(\zeta_2^{(2)} - 2)
r^{-2}\delta^{\alpha\beta}({\bf  r \cdot n})({\bf  r \cdot p}) \nonumber\\
&-& \zeta_2^{(2)}(\zeta_2^{(2)}-2) r^{-2}[(r^\alpha p^\beta + r^\beta
p^\alpha) ({\bf  r \cdot n}) \nonumber\\
&&~\hspace{2.5cm}+(r^\alpha n^\beta + r^\beta n^\alpha)({\bf r\cdot p})]
{\Big ]}\nonumber\\
S^{\alpha\beta}_{j=2,k=-2}({\bf  r})
&=&a_{9,2,-2}r^{\zeta_2^{(2)}}{\Big [}([\zeta_2^{(2)}]^2+5\zeta_2^{(2)}+6)(m^\alpha p^\beta + m^\beta p^\alpha)\nonumber\\
&+&2\zeta_2^{(2)}(\zeta_2^{(2)} - 2)r^{-4} r^\alpha r^\beta
({\bf r\cdot p})({\bf  r \cdot m})\nonumber\\
&-&2\zeta_2^{(2)}(2+\zeta_2^{(2)})r^{-2}[(r^\alpha p^\beta
+ r^\beta p^\alpha)({\bf  r \cdot m})\nonumber\\
&&~\hspace{2.5cm}+(r^\alpha m^\beta + r^\beta m^\alpha)({\bf r\cdot p})]
{\Big ]}\nonumber\\
&+&a_{1,2,-2}r^{\zeta_2^{(2)}}{\Big [}2([\zeta_2^{(2)}]^2-\zeta_2^{(2)}-2)(m^\alpha p^\beta + m^\beta
p^\alpha)\nonumber\\
&+&2\zeta_2^{(2)}(\zeta_2^{(2)} - 2)
r^{-2}\delta^{\alpha\beta}({\bf  r \cdot m})({\bf  r \cdot p}) \nonumber\\
&-&2\zeta_2^{(2)}(\zeta_2^{(2)}-2) r^{-2}[(r^\alpha p^\beta + r^\beta
p^\alpha) ({\bf  r \cdot m})\nonumber\\
&&~\hspace{2.5cm}+(r^\alpha m^\beta + r^\beta m^\alpha)({\bf  r
\cdot p})]{\Big ]}\nonumber\\
S^{\alpha\beta}_{j=2,k=2}({\bf r})
&=&a_{9,2,2}r^{\zeta_2^{(2)}}{\Big [}2([\zeta_2^{(2)}]^2
+5\zeta_2^{(2)}+6)(m^\alpha m^\beta- p^\beta p^\alpha)\nonumber\\
&+& 2\zeta_2^{(2)}(\zeta_2^{(2)} - 2)
r^{-4} r^\alpha r^\beta [({\bf  r \cdot m})^2-({\bf  r \cdot
p})^2]\nonumber\\
&-&2\zeta_2^{(2)}(2+\zeta_2^{(2)})r^{-2}[(r^\alpha m^\beta +
r^\beta m^\alpha)({\bf  r \cdot m})\nonumber\\
&&~\hspace{2.5cm}-(r^\alpha p^\beta + r^\beta p^\alpha)({\bf  r \cdot p})]{\Big ]}
\nonumber\\
&+&a_{1,2,2}r^{\zeta_2^{(2)}}{\Big [}2([\zeta_2^{(2)}]^2-\zeta_2^{(2)}-2)(m^\alpha m^\beta - p^\beta
p^\alpha)\nonumber\\
&+&2\zeta_2^{(2)}(\zeta_2^{(2)} - 2)
r^{-2}\delta^{\alpha\beta}[({\bf  r \cdot m})^2-({\bf  r \cdot
p})^2]\nonumber\\ &-&2\zeta_2^{(2)}(\zeta_2^{(2)}-2) r^{-2}[(r^\alpha
m^\beta + r^\beta m^\alpha) ({\bf  r \cdot m})\nonumber\\
&&~\hspace{2.5cm} - (r^\alpha p^\beta + r^\beta p^\alpha) ({\bf  r
\cdot p})]{\Big ]}.
\end{eqnarray}

Now we wish to use this form to fit for the scaling exponent
$\zeta_2^{(2)}$ in the structure function $S^{33}({\bf r})$ from
data set I where $\alpha=\beta=3$ and the azimuthal angle of ${\bf
r}$ in the geometry is $\phi = \pi/2$.:

\begin{eqnarray}
S^{33}_{j=2,k=-1}(r,\theta,\phi=\pi/2)&=&0\nonumber\\
S^{33}_{j=2,k=1}(r,\theta,\phi=\pi/2)
&=&a_{9,2,1}r^{\zeta_2^{(2)}}[-2\zeta_2^{(2)}(\zeta_2^{(2)}+2)
\sin\theta\cos\theta \nonumber \\
&+& 2\zeta_2^{(2)}(\zeta_2^{(2)}-2)\cos^3\theta\sin\theta]\nonumber \\
S^{33}_{j=2,k=-2}(r,\theta,\phi=\pi/2)&=&0\nonumber \\
S^{33}_{j=2,k=2}(r,\theta,\phi=\pi/2)
&=&a_{9,2,2}r^{\zeta_2^{(2)}}[-2\zeta_2^{(2)}(\zeta_2^{(2)}-2)
\cos^2\theta\sin^2\theta]
\nonumber\\
&+&a_{1,2,2}r^{\zeta_2^{(2)}}[-2\zeta_2^{(2)}(\zeta_2^{(2)}-2)\sin^2\theta].
\end{eqnarray}
\begin{table}
\centerline{\includegraphics[scale=0.8]{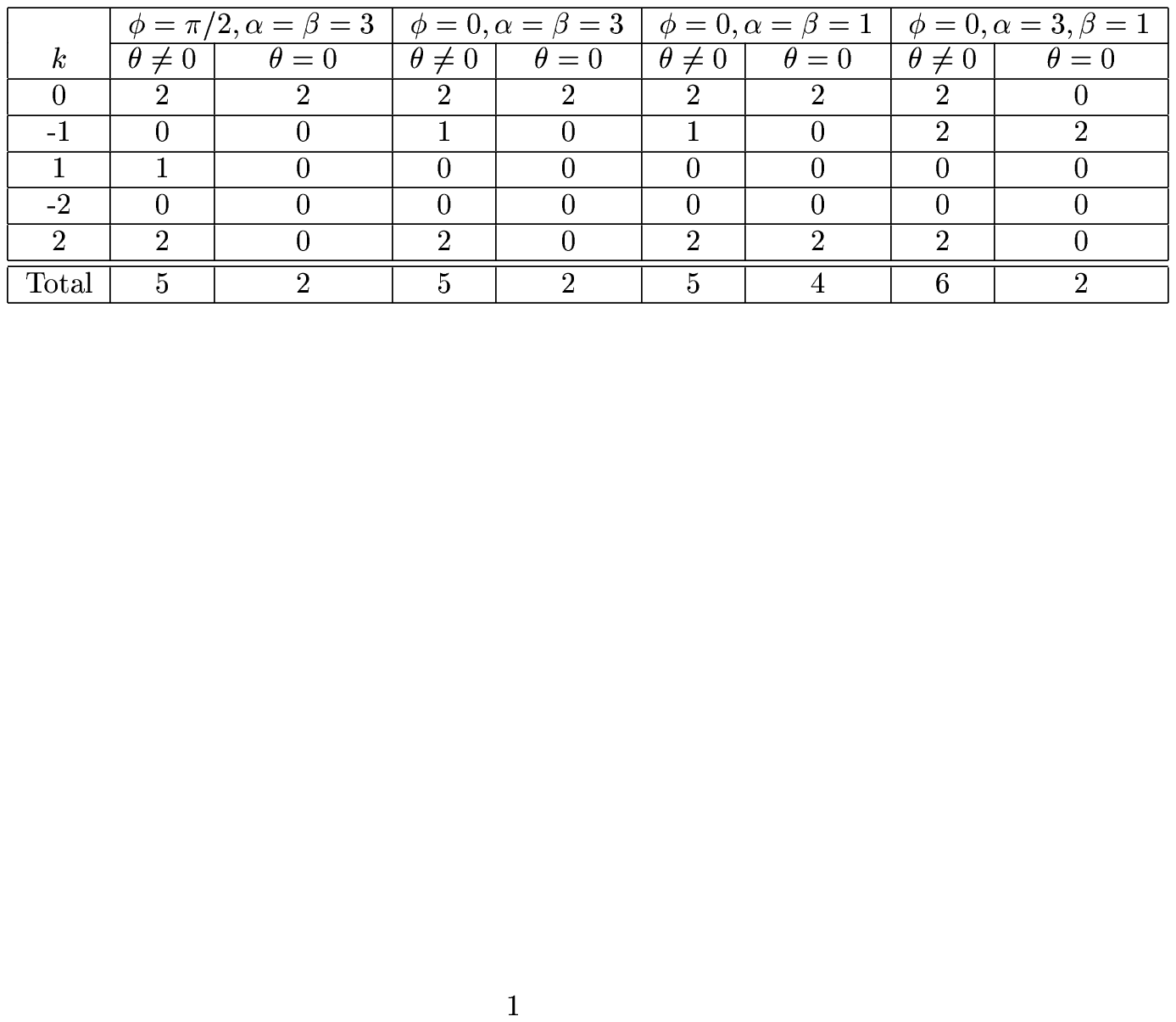}}
\caption{\label{j2_free}The number of free coefficients in the $j=2$ sector for
homogeneous turbulence and for different geometries}
\end{table}
We see that choosing a particular geometry eliminates certain
tensor contributions. In the case of set I we are left with 3
independent coefficients for $m\ne0$, the 2 coefficients from the
$m=0$ contribution (Eq.~\ref{m0}), and the single coefficient from
the isotropic sector \ref{Siso}, giving a total of 6 fit
parameters. The general forms in \ref{ktens} can be used along
with the $k=0$ (axisymmetric) contribution \ref{Siso} to fit to
any second order tensor object. For convenience, Table
\ref{j2_free} shows the number of independent coefficients that a
few different experimental geometries we have will allow in the
$j=2$ sector. It must be kept in mind that these forms are to be
used {\em only} when homogeneity is known to exist. If there is
inhomogeneity, the incompressibility
condition cannot be used to provide constraints in the various parity and
symmetry sectors, and we must in general mix different parity
objects using only the geometry of the experiment itself to
eliminate any terms.
\section{The j=1 component in the inhomogeneous case}
\subsection{Antisymmetric contribution}
We consider the tensor
\begin{equation}
T^{\alpha\beta}({\bf  r}) = <u^\alpha({\bf  x} + {\bf  r}) - u^\alpha({\bf
x}))
(u^\beta({\bf  x} + {\bf  r}) + u^\beta({\bf  x}))>.
 \end{equation}
This object is trivially zero for $\alpha=\beta$. In our
experimental setup, we measure at points separated in the shear
direction and therefore have inhomogeneity which makes the object
of mixed parity and symmetry.  We cannot apply the
incompressibility condition in same parity/symmetry sectors as
before to provide constraints. We must in general use all 7
irreducible tensor forms. This would mean fitting for $7 \times 3
= 21$ independent coefficients plus 1 exponent $\zeta_2^{(1)}$ in
the anisotropic sector, together with 2 coefficients in the
isotropic sector. In order to pare down the number of parameters
we are fitting for, we look at the antisymmetric part of
$T^{\alpha\beta}({\bf  r})$,
\begin{equation}
{\widetilde T}^{\alpha\beta}({\bf  r}) = {T^{\alpha\beta}({\bf
r}) - T^{\beta\alpha}({\bf  r}) \over 2} = \langle u^\alpha({\bf
x})u^\beta({\bf  x} + {\bf  r})\rangle - \langle u^\beta({\bf
x})u^\alpha({\bf  x} + {\bf  r})\rangle, \end{equation} which will
only have contributions from the antisymmetric $j=1$ basis
tensors. These are: \\
$\bullet$ Antisymmetric, odd parity
\begin{equation}
B_{3,1,m}^{\alpha\beta}
= r^{-1}[r^\alpha\partial^\beta- r^\beta\partial^\alpha] rY_{1,m}(\bf  {\hat r})
\end{equation}
$\bullet$ Antisymmetric, even parity
\begin{eqnarray}
B_{4,1,m}^{\alpha\beta} &=&
r^{-2}\epsilon^{\alpha\beta\mu}r_\mu r Y_{1,m}(\bf  {\hat r}) \nonumber\\
B_{2,1,m}^{\alpha\beta} &=&
r^{-2}\epsilon^{\alpha\beta\mu}\partial_\mu r Y_{1,m}(\bf  {\hat
r}).
\end{eqnarray}

As with the $j=2$ case we form a real basis $r {\tilde
Y}_{1,k}(\bf {\hat r})$ from the (in general) complex $r
Y_{1,m}(\bf {\hat r})$ in order to obtain real coefficients in our
fits:
\begin{eqnarray}
r {\tilde Y}_{1,k=0}({\bf{\hat r}}) &=& r Y_{1,0}({\bf{\hat r}})=
r\cos\theta = r_3 \nonumber\\
r {\tilde Y}_{1,k=1}({\bf{\hat r}}) &=& r {Y_{1,1}({\bf{\hat r}})+Y_{1,1}
({\bf{\hat r}}) \over 2i } \nonumber \\
&=&r\sin\theta\sin\phi = r_2\nonumber\\
r {\tilde Y}_{1,k=-1}(\bf  {\hat r})
&=& r{Y_{1,-1}({\bf{\hat r}}) - Y_{1,1}({\bf{\hat r}})\over 2}\nonumber \\
&=&r\sin\theta\cos\phi = r_1 \nonumber\\
\end{eqnarray}
And the final forms are
\begin{eqnarray}
B_{3,1,0}^{\alpha\beta}({\bf{\hat r}})
&=& r^{-1}[r^\alpha n^\beta- r^\beta n^\alpha] \nonumber\\
B_{4,1,0}^{\alpha\beta}({\bf{\hat r}})
&=&r^{-2}\epsilon^{\alpha\beta\mu}r_\mu ({\bf  r.n}) \nonumber\\
B_{2,1,0}^{\alpha\beta}({\bf{\hat r}})
&=& r^{-2}\epsilon^{\alpha\beta\mu}n_\mu \nonumber\\
B_{3,1,1}^{\alpha\beta}({\bf{\hat r}})
&=& r^{-1}[r^\alpha p^\beta- r^\beta p^\alpha] \nonumber\\
B_{4,1,1}^{\alpha\beta}({\bf{\hat r}})
&=&r^{-2}\epsilon^{\alpha\beta\mu}r_\mu ({\bf  r.p}) \nonumber\\
B_{2,1,1}^{\alpha\beta}({\bf{\hat r}})
&=& r^{-2}\epsilon^{\alpha\beta\mu}p_\mu \nonumber\\
B_{3,1,-1}^{\alpha\beta}({\bf{\hat r}})
&=& r^{-1}[r^\alpha m^\beta- r^\beta m^\alpha] \nonumber\\
B_{4,1,-1}^{\alpha\beta}({\bf{\hat r}})
&=&r^{-2}\epsilon^{\alpha\beta\mu}r_\mu ({\bf  r.m}) \nonumber\\
B_{2,1,-1}^{\alpha\beta}({\bf{\hat r}}) &=&
r^{-2}\epsilon^{\alpha\beta\mu}m_\mu. \end{eqnarray}

We now have 9 independent terms and cannot apply incompressibility
in order to reduce them further. Instead, we use the geometrical
constraints of the experiment to do this:

$\bullet$ $\phi = 0$ (vertical separation), $\alpha = 3, \beta = 1$
\begin{eqnarray}
B_{3,1,0}^{31}(r,\theta,\phi=0) = -\sin\theta \nonumber
\\ B_{2,1,1}^{31}(r,\theta,\phi=0) = 1 \nonumber\\
B_{3,1,-1}^{31}(r,\theta,\phi=0) = \cos\theta.
\end{eqnarray}

There are no contributions from the reflection-symmetric terms in
the $j=0$ isotropic sector since these are symmetric in the
indices. The helicity term in $j=0$ also does not contribute
because of the geometry. So, to lowest order, we have
\begin{eqnarray}
{\widetilde T}^{31}({\bf  r})
&=& {\widetilde T}_{j=1}^{31}({\bf  r}) \nonumber
\\ &=& a_{3,1,0}r^{\zeta_2^{(1)}}(-\sin\theta) + a_{2,1,1}r^{\zeta_2^{(1)}} +
a_{3,1,-1}r^{\zeta_2^{(1)}}\cos\theta.
\label{T31_final}
\end{eqnarray}
As always, we have written the scale dependent prefactors
$a_{qjm}(r)$ as having a power law dependence
$a_{qjm}r^{\zeta_2^{(1)}}$. We have 3 unknown independent
coefficients and 1 unknown exponent to fit for in the data.

\subsection{Symmetric contribution}

We consider the structure function
\begin{equation}
S^{\alpha\beta}({\bf  r}) = <(u^\alpha({\bf  x} + {\bf  r}) -
u^\alpha({\bf x})) (u^\beta({\bf  x} + {\bf  r}) - u^\beta({\bf
x}))> \end{equation} in the case where we have homogeneous flow.
This object is symmetric in the indices by construction. It is
easily seen that homogeneity implies even parity in r:
\begin{eqnarray}
    S^{\alpha\beta}({\bf  r}) &=& S^{\beta\alpha}({\bf  r})\nonumber \\
S^{\alpha\beta}({\bf  -r}) &=& S^{\alpha\beta}({\bf  r}).
\end{eqnarray}
We reason that this object cannot exhibit a $j=1$
contribution from the $SO(3)$ representation in the following. 
Homogeneity allows us to use the incompressibility
condition
\begin{eqnarray}
    \partial_\alpha S^{\alpha\beta} &=& 0 \nonumber\\
    \partial_\beta S^{\alpha\beta} &=& 0
\end{eqnarray}
separately on the basis tensors of a given parity and symmetry in
order to give relationships between their coefficients. For even
parity, symmetric case we have for general $j \geq 2$ just two
basis tensors and they must occur in some linear combination with
incompressibility providing a constraint between the two
coefficients. However, for $j=1$ we only have one such tensor in
the even parity, symmetric group. Therefore, by incompressibility,
its coefficient must vanish. Consequently, we cannot have a $j=1$
contribution for the even parity (homogeneous), symmetric
structure function.

Now, we consider the case of an experiment when ${\bf r}$ has some
component in the inhomogeneous direction. Now, it is no longer
true that $S^{\alpha\beta}({\bf r})$ is of even parity. Moreover
it is not possible to use incompressibility as above to exclude
the existence of a $j=1$ contribution. We must look at all $j=1$
basis tensors that are symmetric, but not confined to even
parity. These are: \\
$\bullet$ Odd parity, symmetric
\begin{eqnarray}
B_{1,1,k}^{\alpha\beta}({{\bf{\hat r}}})
&\equiv& r^{-1}\delta^{\alpha\beta}r {\tilde Y}_{1k}({\bf  {\hat
r}})\nonumber\\ B_{7,1,k}^{\alpha\beta}({{\bf{\hat r}}})
&\equiv& r^{-1}[r^\alpha \partial^\beta + r^\beta \partial^\alpha] r{\tilde
Y}_{1k}({\bf{\hat r}}) \nonumber \\ B_{9,1,k}^{\alpha\beta}({{\bf{\hat r}}})
&\equiv& r^{-3}r^\alpha r^\beta r {\tilde Y}_{1k}({{\bf{\hat r}}})\nonumber
\\ B_{5,1,k}^{\alpha\beta}({{\bf{\hat r}}})
&\equiv& r \partial^\alpha \partial^\beta r{\tilde Y}_{1k}({{\bf{\hat r}}})
\equiv 0
\end{eqnarray}
$\bullet$ Even parity, symmetric
\begin{eqnarray}
B_{8,1,k}^{\alpha\beta}({{\bf{\hat r}}})
&\equiv& r^{-2}[r^\alpha \epsilon^{\beta\mu\nu} r_\mu \partial_\nu +
r^\beta\epsilon^{\alpha\mu\nu} r_\mu \partial_\nu] r{\tilde Y}_{1k}
({\bf{\hat r}})\nonumber\\
B_{6,1,k}^{\alpha\beta}({{\bf{\hat r}}}) &\equiv&
[\epsilon^{\beta\mu\nu} r_\mu \partial_\nu \partial_\alpha +
\epsilon^{\beta\mu\nu} r_\mu \partial_\nu \partial_\beta] r{\tilde
Y}_{1k}({\bf{\hat r}}) \equiv 0.
\end{eqnarray}
We use the real basis of $r^{-1}{\tilde Y}_{1k}({{\bf{\hat r}}})$
which are formed from the $r^{-1}Y_{1m}({{\bf{\hat r}}})$. Both
$B_{5,1,k}^{\alpha\beta}({{\bf{\hat r}}})$ and
$B_{6,1,k}^{\alpha\beta}({\bf {\hat r}})$ vanish because of the
taking of the double derivative of an object of single power in
$r$. We thus have 4 different contributions to symmetric $j=1$ and
each of these is of 3 dimensions $(k= -1,0,1)$ giving in general
12 terms in all: \begin{eqnarray}
B_{1,1,0}^{\alpha\beta}({{\bf{\hat r}}})
&=& r^{-1}\delta^{\alpha\beta}({\bf  r \cdot n})\nonumber\\
B_{7,1,0}^{\alpha\beta}({{\bf{\hat r}}})
&=& r^{-1}[r^\alpha n^\beta + r^\beta n^\alpha] \nonumber\\
B_{9,1,0}^{\alpha\beta}({{\bf{\hat r}}})
&=&r^{-3}r^\alpha r^\beta ({\bf  r \cdot n})\nonumber\\
B_{8,1,0}^{\alpha\beta}({{\bf{\hat r}}})
&\equiv& r^{-2}[(r^\alpha m^\beta + r^\beta m^\alpha)({\bf  r \cdot p})
-(r^\alpha p^\beta + r^\beta p^\alpha)({\bf  r \cdot m})]\nonumber \\
B_{1,1,1}^{\alpha\beta}({{\bf{\hat r}}})
&=& r^{-1}\delta^{\alpha\beta}({\bf  r \cdot p})\nonumber\\
B_{7,1,1}^{\alpha\beta}({{\bf{\hat r}}})
&=& r^{-1}[r^\alpha p^\beta + r^\beta p^\alpha] \nonumber\\
B_{9,1,1}^{\alpha\beta}({{\bf{\hat r}}})
&=&r^{-3}r^\alpha r^\beta ({\bf  r \cdot p}) \nonumber\\
B_{8,1,1}^{\alpha\beta}({{\bf{\hat r}}})
&\equiv& r^{-2}[(r^\alpha m^\beta + r^\beta m^\alpha)({\bf  r \cdot n})
-(r^\alpha n^\beta + r^\beta n^\alpha)({\bf  r \cdot m})] \nonumber \\
B_{1,1,-1}^{\alpha\beta}({{\bf{\hat r}}}) &=&
r^{-1}\delta^{\alpha\beta}({\bf  r \cdot m})\nonumber\\
B_{7,1,-1}^{\alpha\beta}({{\bf{\hat r}}}) &=& r^{-1}[r^\alpha m^\beta +
r^\beta m^\alpha] \nonumber\\ B_{9,1,-1}^{\alpha\beta}({{\bf{\hat r}}})
&=&r^{-3}r^\alpha r^\beta ({\bf  r \cdot m}) \\
B_{8,1,-1}^{\alpha\beta}({{\bf{\hat r}}}) &\equiv&
r^{-2}[(r^\alpha p^\beta + r^\beta p^\alpha)({\bf  r \cdot n})
-(r^\alpha n^\beta + r^\beta n^\alpha)({\bf  r \cdot p})].
\nonumber
\end{eqnarray}
These are all the possible $j=1$ contributions to the symmetric, mixed
parity (inhomogeneous) structure function.
\begin{table}
\centerline{\begin{tabular}{|c|c|c|c|c|c|c|}\hline
&\multicolumn{2}{c|}{$\phi= 0,\alpha=\beta=3$} & \multicolumn{2}{c|}{$\phi
= 0,\alpha=\beta=1$} & \multicolumn{2}{c|}{$\phi = 0,\alpha=3,\beta=1$}\\
\cline {2-7} $k$&$\theta \ne 0$ &$\theta = 0$ & $\theta \ne 0$ &$ \theta =
0$ & $\theta \ne 0$ & $\theta = 0$ \\ \hline
0 & 3 & 3 & 2 & 1 & 2 & 0 \\ \hline
1 & 1 & 0 & 1 & 0 & 0 & 0 \\ \hline
-1 & 2 & 0 & 3 & 0 & 2 & 1 \\ \hline \hline
Total & 6 & 3 & 6 & 1 & 4 & 1\\\hline
\end{tabular}}
\caption{\label{j1_coeffs}The number of free coefficients in the symmetric $j=1$ sector for
inhomogeneous turbulence and for different geometries.} \end{table}

For our experimental setup II, we want to analyze the
inhomogeneous structure function in the case $\alpha = \beta = 3$,
and azimuthal angle $\phi=0$ (which corresponds to vertical
separation). For that case, we obtain the basis tensors to be
\begin{eqnarray}
B_{1,1,0}^{33}(\theta) &=& \cos\theta \nonumber\\ B_{7,1,0}^{33}(\theta)
&=& 2\cos\theta \nonumber\\ B_{9,1,0}^{33}(\theta) &=& \cos^3\theta
\nonumber\\ B_{8,1,1}^{33}(\theta) &=& -2\cos\theta\sin\theta \nonumber\\
B_{1,1,-1}^{33}(\theta) &=& \sin\theta \nonumber\\
B_{9,1,-1}^{33}(\theta) &=& \cos^2\theta\sin\theta. \end{eqnarray}

Table \ref{j1_coeffs} gives the number of free coefficients in the symmetric $j=1$ sector
in the fit to the inhomogeneous structure function for various geometrical
configurations.

\section{Tests of the robustness of the interpolation formula}
In order to test the robustness of the interpolation formula 
Eq.~\ref{interp}, we
performed the following additional calculations. We considered the
data from the probe at the height of 0.54 m. For each order $n$ of the
structure function, we defined a `window' of data extending over two
decades of the separation scale, $r$. We first placed the lower edge
of the window well inside the dissipation range and fit the
interpolation formula to the data in the first window. We then moved
the lower edge of the window by half a decade and fit the formula to
the data in the next window. In this manner, we proceeded until the
upper edge of the last window corresponded to the largest value of
$r$. The entire range of $r$ yields five windows. We thus obtained
five values of the parameter $C_n$ and calculate the scaling exponent
$\zeta_n^{(2)}=n - 2C_n$ in each case, giving some indication of the
robustness of our result.
\begin{table}[!ht]
\begin{tabular}{|c|c|c|c|c|c|}\hline
$C_2$&$0.35\pm0.1$&$0.35\pm0.05$&$0.39\pm0.02$&$0.38\pm0.05$
&$0.38\pm0.07$\\\hline
$\zeta_2^{(2)}$&$1.31\pm0.2$&$1.30\pm0.10$&$1.21\pm0.04$&$1.24\pm0.10$&
$1.23\pm0.14$\\\hline
\end{tabular}
\caption{Second order: $\zeta_2^{(2)} = 1.25 \pm 0.05$.\label{test_2}}
\end{table}
\begin{table}[!ht]
\begin{tabular}{|c|c|c|c|c|c|}\hline
$C_3$&$0.99\pm0.03$&$0.95\pm0.04$&$0.88\pm0.07$&$0.91\pm0.04$&
$0.96\pm0.08$\\\hline
$\zeta_3^{(2)}$&$1.01\pm0.06$&$1.10\pm0.08$&$1.3\pm0.14$&$1.2\pm0.08$&
$1.1\pm0.16$\\\hline
\end{tabular}
\caption{Third order: $\zeta_3^{(2)} = 1.14 \pm 0.11$.\label{test_3}}
\end{table}
\begin{table}[!ht]
\begin{tabular}{|c|c|c|c|c|c|}\hline
$C_4$&$1.21\pm0.07$&$1.12\pm0.09$&$1.15\pm0.03$&$1.29\pm0.1$
&$1.21\pm0.08$\\\hline
$\zeta_4^{(2)}$&$1.58\pm0.14$&$1.76\pm0.18$&$1.7\pm0.06
$&$1.42\pm0.2$&$1.58\pm0.16$ \\\hline
\end{tabular}
\caption{Fourth order: $\zeta_4^{(2)} = 1.61 \pm 0.13$.\label{test_4}}
\end{table}

Tables \ref{test_2}-\ref{test_4} present the results of performing
these checks on structure functions of the second, third and
fourth order.  The mean and standard deviation of the exponent
values are given in the caption for each table. It is found that
the mean value in each case is in close agreement to the value of
the exponents presented in the main text which were obtained by a
fit to the entire range of data. This gives us greater confidence
in the use of the interpolation formula.

\bibliography{thesis}       
\bibliographystyle{plain}
\end{document}